\documentclass[twocolumn,showpacs,preprintnumbers,amsmath,amssymb,prb,aps]{revtex4}
\usepackage{latexsym}
\usepackage{graphicx}
\usepackage{graphics}
\usepackage{bm}
\usepackage{lscape}
\usepackage{subfigure}
\usepackage{xcolor}

\begin{document}

\title [] {Stoichiometric and off-stoichiometric \\ 
full Heusler $\mathbf {Fe_2V_{1-x}W_xAl} $ thermoelectric systems
}

\author{B. Hinterleitner$^{1,2}$, P. Fuchs$^{1}$, J. Rehak$^{1}$,  
S. Steiner$^{1,5}$, M. Kishimoto$^{3}$, R. Moser$^{4}$,
R. Podloucky$^{4}$, E. Bauer$^{1,2}$, }

\affiliation{$^1$Institute of Solid State Physics, Technische Universit\"at Wien, A-1040 Wien, Austria}

\affiliation{$^2$Christian Doppler Laboratory for Thermoelectricity, Technische Universit\"at Wien, A-1040 Wien, Austria}

\affiliation{$^3$Tokyo Metropolitan University}

\affiliation{$^4$Institute of Applied Physics, Technische Universit\"at Wien, A-1040 Wien, Austria}

\affiliation{$^5$Central Europian Institute of Technology, 
Brno University of Technology, Purkynova 123, 61200 Brno, Czech Republik}


\begin{abstract}

A series of full-Heusler alloys, $\rm Fe_2V_{1-x}W_xAl$, $0 \leq x \leq 0.2$, 
was prepared, characterized and relevant physical properties 
to account for the thermoelectric performance were studied in a wide temperature range.
Additionally, off-stoichiometric samples with similar compositions have been
included, and a 10~\% improvement of the thermoelectric figure of merit
was obtained. The V/W substitution causes i) a change of the main carrier type,
from holes to electrons as evidenced from Seebeck and Hall measurements and ii) a
substantial reduction of the lattice thermal conductivity due to a creation of lattice disorder 
by means of a distinct different mass and metallic radius upon the V/W substitution. 
Moreover $ZT$ values above 0.2 have been obtained. 
A microscopic understanding of the experimental data observed is revealed from
ab-initio calculations of the electronic and phononic structure.
This series of alloys constitutes the basis for thin film systems, which have recently been
found to exhibit $ZT$ values beyond those reported so far in literature.


\end{abstract}

\pacs{abc}
\maketitle

\section{Introduction}

Half-Heusler XYZ and full-Heusler systems, $\rm X_2YZ$ have attracted growing interest 
in many functional materials related fields, like in spintronics, 
optoelectronic, superconductivity,
shape memory or thermoelectricity. For the latter, a variety of promising properties
like the mechanical and chemical stability and for several members of XYZ systems a 
thermoelectric performance with good prospects have been found. Here, X and Y are in general 
transition metal elements and Z is a main group element like aluminium. While XYZ systems
consist of 3 $fcc$ sublattices, $\rm X_2YZ$ systems are composed of four $fcc$ sublattices.
Since Heusler systems are prone for anti-side occupations, the cubic crystal structures might 
change with respect to its space groups and $fcc$ based crystals might transform to $bcc$ type
materials. Recent reviews summarise  many of the 
above indicted aspects \cite{Graf2011,felser2015heusler,Felser2016}. 

Full-Heusler compounds are known for its closeness to an insulating state, i.e, the proximity 
of the Fermi energy to a gap in the electronic density of states (eDOS). 
Specifically, the Fermi energy
$E_F$ of $\rm Fe_2VAl$ is located very slightly 
below a pseudo-gap in the eDOS (see e.g., Ref. \onlinecite{Weinert_1998}). 
Consequently, holes are the main charge carriers 
and positive Seebeck and Hall values can be obtained 
experimentally \cite{Nishino_2006}. Small changes of the electron 
concentration by electron or hole doping is expected
to shift either $E_F$ towards the conduction band or 
more deeply into the valence band, resulting 
in electron or hole dominated transport, as 
one of the pre-requisites of thermoelectric materials.

$\rm Fe_2VAl$ offers the possibility to exchange elements on all 
lattice sites, i.e., on the Fe-site (8c)
on the V site (4a) and the Al site (4b) of the $\rm Cu_2MnAl$ structure type. 
Appropriate substitutions
may generate two beneficial features: i) an intended modification and tuning of the eDOS and
ii) disorder in the crystal lattice, expected to distinctly decrease the 
lattice thermal conductivity of the material.

A variety of studies regarding substitutions, doping, or off-stoichiometric sample preparation 
has already been carried out previously and reported in literature
\cite{Nishino_2006,Lue_2007,Mikami_2008,
Mikami_2009,Mikami_2012,Renard_2014,Mikami_2015,Wei_2015,Yamani_2015}. 
Most successful studies with respect to the thermoelectric performance,
expressed by the dimensionless figure of merit $ZT = S^2 / (\rho \lambda)$, where $T$ is the 
absolute temperature, $S$ is the Seebeck effect, $\rho$  the electrical resistivity and
$\lambda$ is the thermal conductivity, have been undertaken by 
substituting Al by Si or Ge, V/W and V/Mo.
$ZT$ values of up to about 0.2 have been obtained. Severe plastic deformation 
in $\rm Fe_2VAl_{0.95}Ta_{0.05}$ caused a further increase of$ZT$ to $ZT = 0.3$ \cite{Masuda_2018}

The aim of the present work is to demonstrate such an advantageous tunability of 
$\rm Fe_2VAl$ by substituting V by W and, in addition, by making this substitution 
off-stoichiometric, i.e., changing $\rm Fe_{50}(V,W)_{25}Al_{25}$  
towards $\rm Fe_{50}(V,W)_{24}Al_{26}$.
In this study, experimentally derived results, obtained on samples up to 20~\% exchange of V by W,
are presented and analyzed and these findings are compared 
and confronted with first principles DFT calculations, allowing to draw some microscopic picture
of this series of alloys. 
Additionally, high temperature and high magnetic field Hall measurements, 
which have been carried out for this study, 
allow to conclude about the predominant charge carriers and their 
respective mobilities, as well as to derive their temperatures dependencies 
at high temperatures. Such a study is lacking in the in the literature up to now.

This series of alloys constitutes the basis for magnetron sputtered 
thin film systems, which have recently been
found to exhibit $ZT$ values beyond those reported so far in literature \cite{nature_2019}. 
A deeper understanding of relevant mechanisms and interactions of the electronic and 
thermal transport is thus of significance.

\section{Experimental}
	
$\rm Fe_2V_{1-x}W_xAl$ alloys and off-stoichiometric variants
(about 3.5 gram each) were
prepared by high frequency melting of high purity starting materials 
on a water-cooled copper hearth
under argon atmosphere. 
To ensure homogeneity of the samples, the melting process was repeated 
several times. Subsequently, all alloys were sealed in quartz
tubes and annealed at 900$^{\circ}$C for one week.

X-ray powder diffraction data at room temperature 
from as-cast and annealed alloys were taken on a Siemens D5000 diffractometer
 employing $\rm CuK_{\alpha1}$ radiation 
$(10<2\Theta<100^{\circ})$. Refinement of the crystal
structures was performed with the program PowderCell. 

Measurements of the various physical properties were
carried out with a series of standard techniques \cite{Bauer_2002}.
A commerically available set-up (ULVAC, ZEM3) was used to study the 
electrical resistivity and the Seebeck coeffient above room temperature.
The Hall resistance was derived above room temperature
using the Van der Pauw technique and employing a LakeShore 370 a.c. resistance bridge.
Fields up to 10\,T were generated by a superconducting magnet (Cryogenic, London) cooled
by a closed cycle refrigerator.

\section{Computational aspects\label{sec:comp}}

For the calculation of the density functional theory (DFT) results the Vienna {\em ab
initio} simulation package (VASP)\cite{Kresse_1996cms,Kresse_1996prb} with the
projector augmented wave potential (PAW)\cite{Blochl_1994prb,Kresse_1999prb}
construction was applied.  For approximating the exchange-correlation functional
we made use of the generalized gradient approximation of
Ref.~\onlinecite{Perdew_1992prb} (GGA) which includes the approach of Vosko, Wilk and
Nusair \cite{Vosko_1980} for spin polarization.

The potentials were constructed in such a way that for V and Fe the
semi-core 3s$^2$ and 3p$^6$ states were treated as valence states resulting in
13 and 16  valence states, respectively. For Al the three  valence states 3s$^2$
and 3p$^1$ were considered. An energy cutoff for the plane wave basis of 600 eV was chosen.
Because a large number of $\mathbf{k}$-points is in particularly needed for the
calculation of thermoelectric properties a grid corresponding to the VASP
parameter KSPACING=0.04  was constructed for all of the calculations. For the
4-atom fcc-unit cell this choice results in 48$^3$ $\mathbf{k}$-points in total
amounting to 2769 symmetry reduced $\mathbf{k}$-points. For the derivation of
local properties such as l-projected densities of states
spheres of radius of 2.0 \AA\, for Fe,
2.3 \AA\, for V,  and 2.65 \AA\, for Al were circumscribed for each corresponding atomic
position.  

For enhancement of exchange-correlation interactions  LDA+U  
(in our case rather GGA+U) methods for the 3d wavefunctions of Fe and occasionally also for
the V 3d states were applied.
Most of LDA+U  calculations were done choosing the VASP
parameter LDAUTYPE=2 which makes use of the simplified rotationally invariant approach
of Dudarev \textit{et al.} \cite{Dudarev_1998}.

The thermoelectric properties were calculated within Boltzmann's transport theory
applying an adapted version of the package BoltzTrap \cite{BoltzTrap}. The
rather large number of $\mathbf{k}$-points was refined by choosing the
interpolation parameter LPFAC=13, because the results are very sensitive due the
fineness of the mesh.

For vibrational properties within the harmonic approximation the 
direct force-constant-method as implemented in the program package 
\textit{f}PHON was used \cite{Alfe2009,DReith}.  For deriving the force constants density
functional perturbation theory  calculations as implemented in 
VASP were performed. Atomic positions were optimized until the 
residual forces were less than $2\times10^{-5}$\,eV/\AA{}. For anharmonic 
contributions a quasi-harmonic approach was applied \cite{Zhang,RMoser}.

\section{Results and Discussion}

\begin{figure}[tbh]
\begin{center}
		\includegraphics[width=0.45\textwidth]{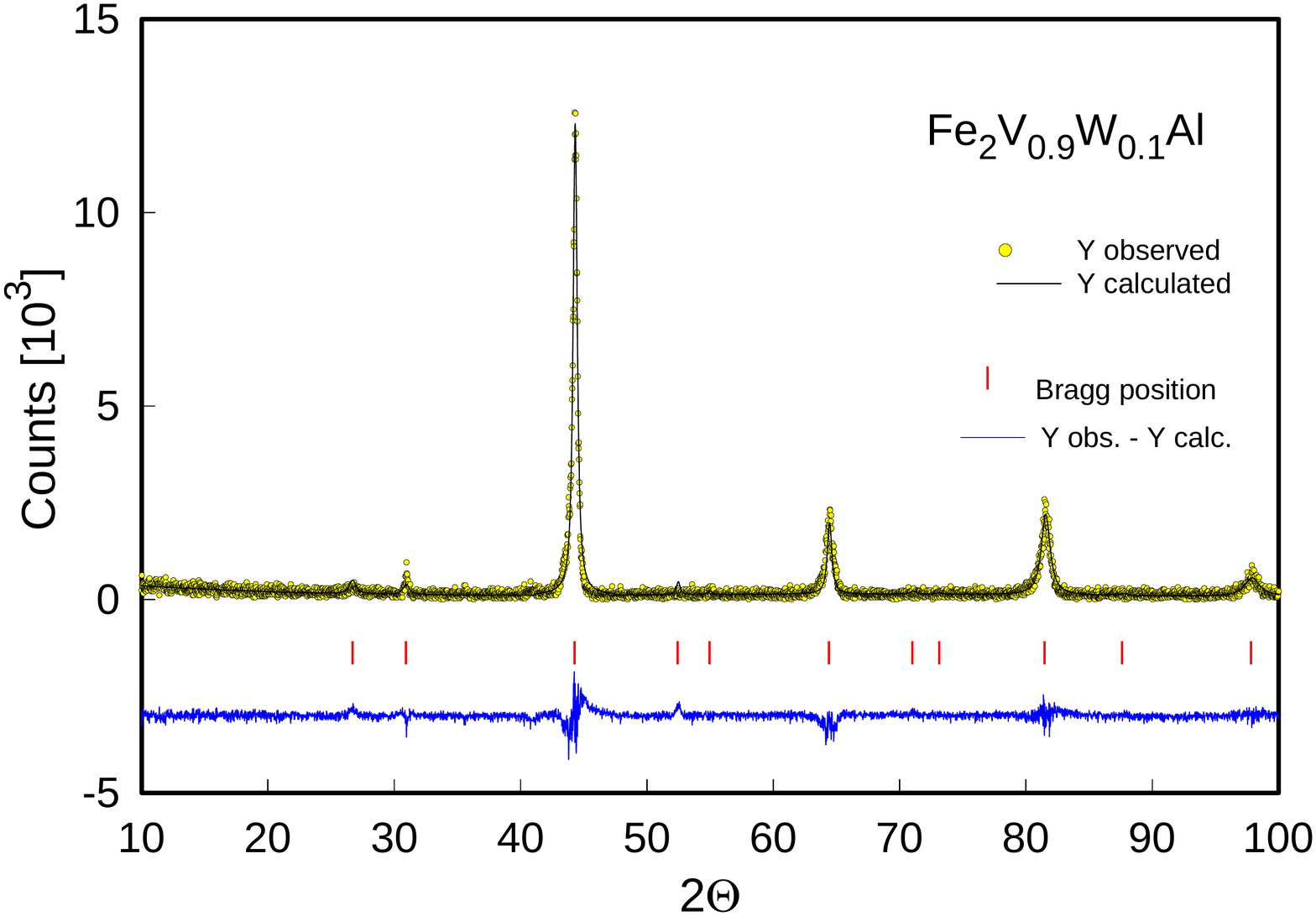}
		\includegraphics[width=0.45\textwidth]{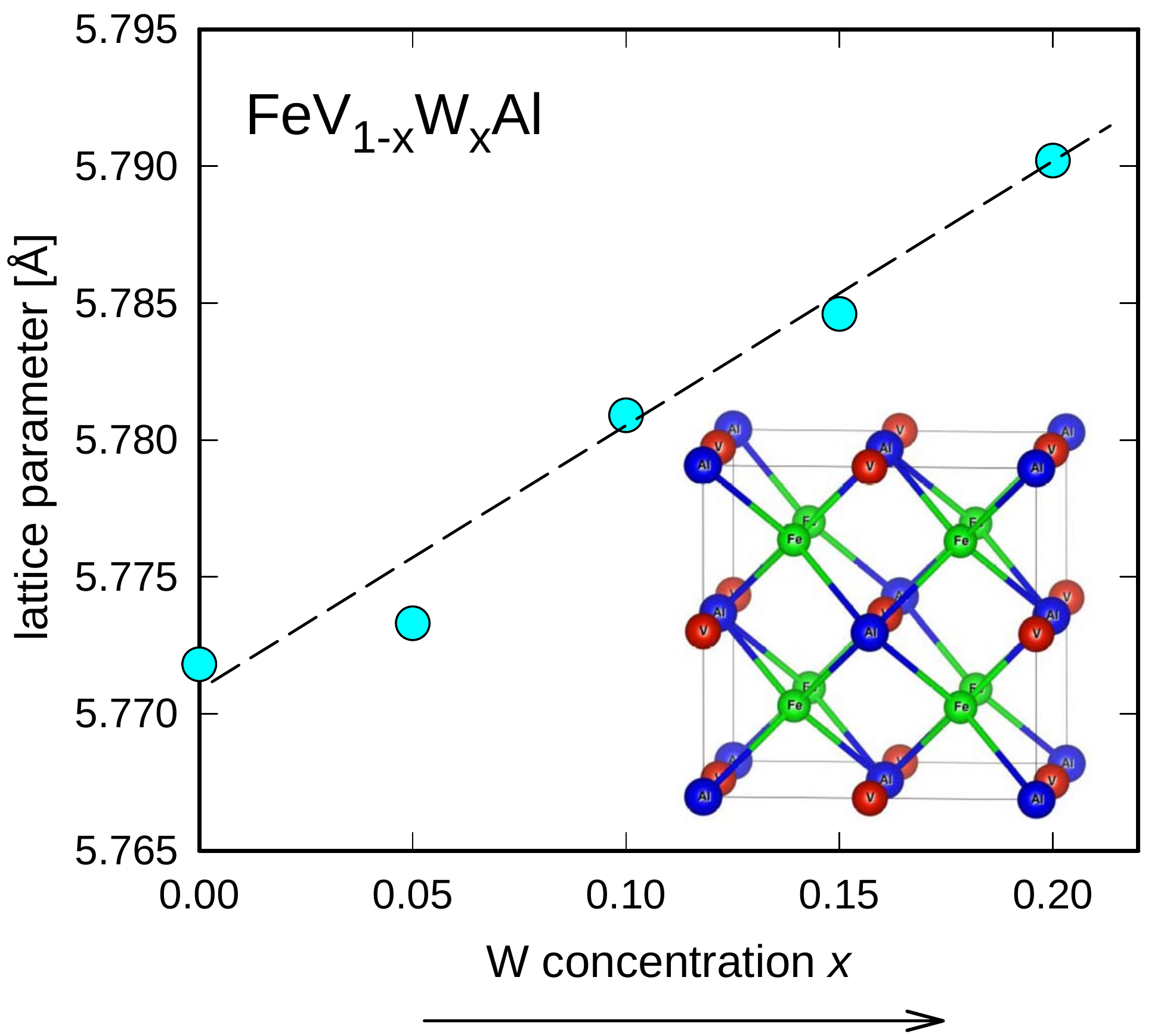}		
\end{center}
	\caption{(Color online) Upper panel: X-ray diffraction 
	pattern for $\rm FeV_{0.9}W_{0.1}Al$, (filled circles),
	together with results from a Rietveld refinement (solid line), 
	the difference between experimental and model data
	(lower part of the figure) and respective Bragg positions. 
	Lower panel: Concentration dependence of the lattice
	parameter $a$ of $\rm FeV_{1-x}W_{x}Al$ taken at room temperature. 
	The inset is a sketch of the crystal 
	structure of $\rm Fe_2VAl$. 
	}
	\label{xray}
\end{figure}

Figure \ref{xray}, upper panel, shows  an example of the x-ray diffraction pattern taken for the 
sample $x = 0.1$. The respective Rietveld refinement is shown in 
addition (solid line), together with 
the respective Bragg positions (vertical lines) and the difference 
between the experimental and the fit data.
Since besides the positions of the expected Bragg peaks no intensity 
is found in the respective X-ray data, good phase purity is evident.

\begin{figure}[h]
\begin{center}
		\includegraphics[width=0.45\textwidth]{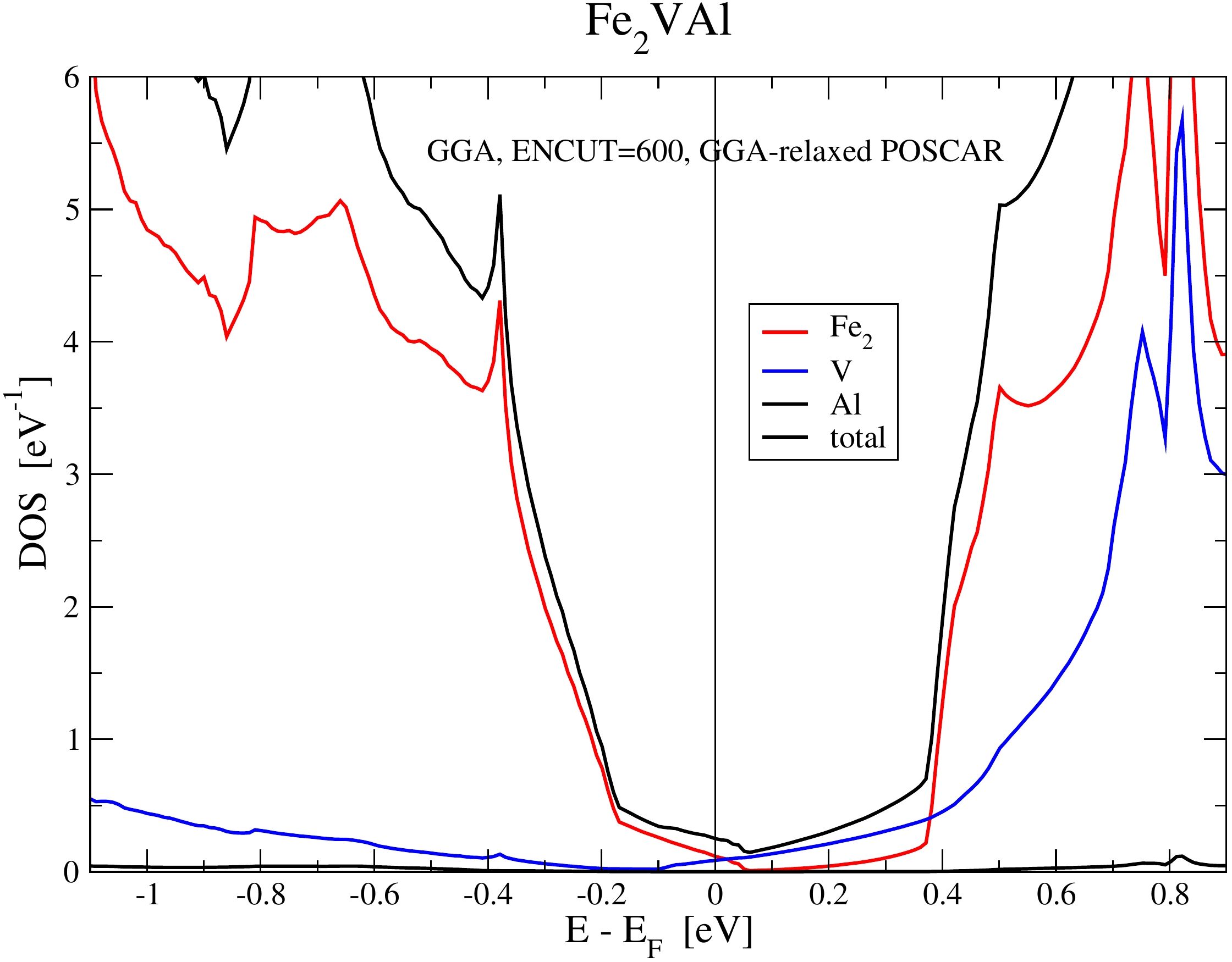}
		\includegraphics[width=0.45\textwidth]{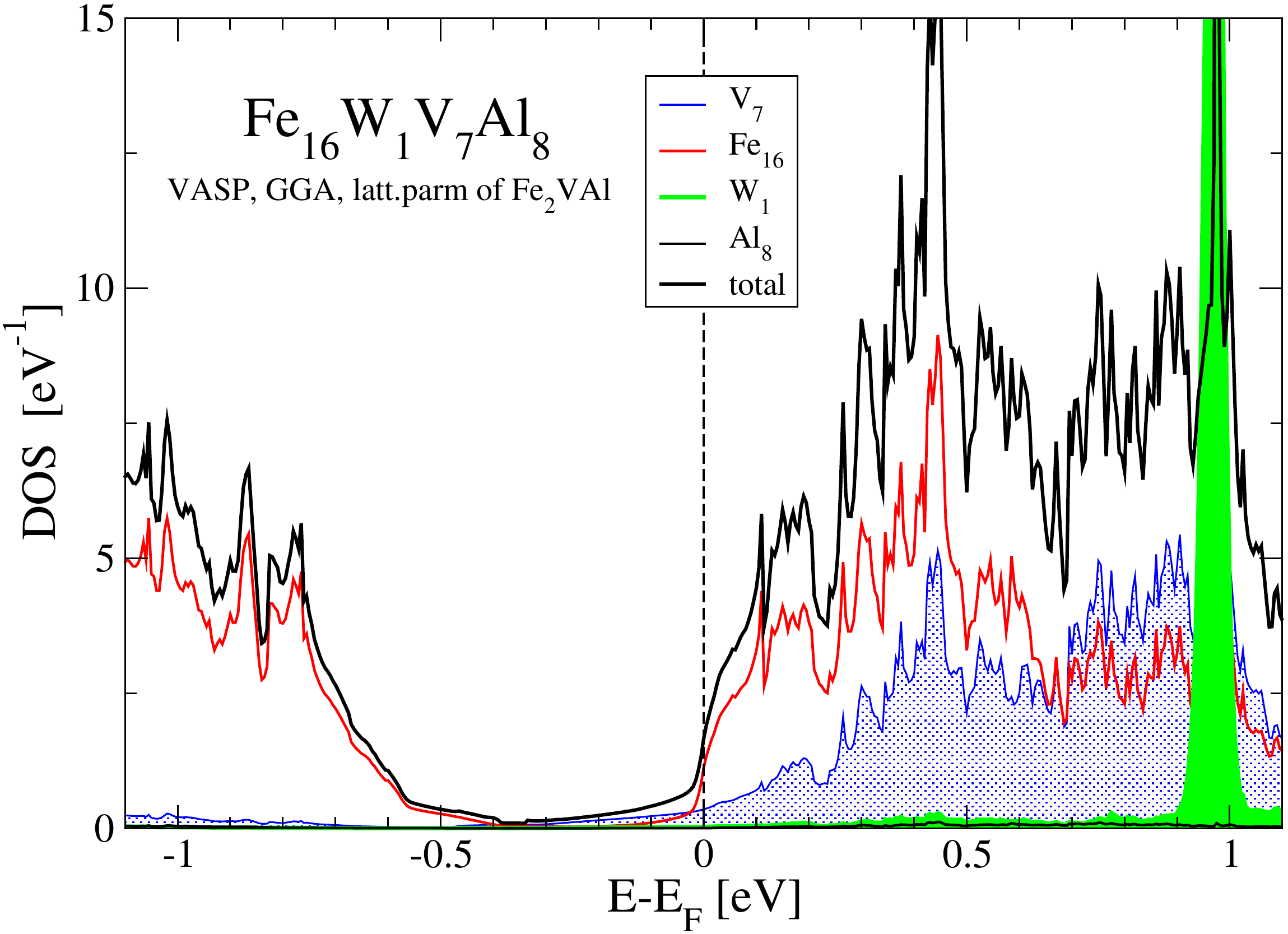}
		\includegraphics[width=0.45\textwidth]{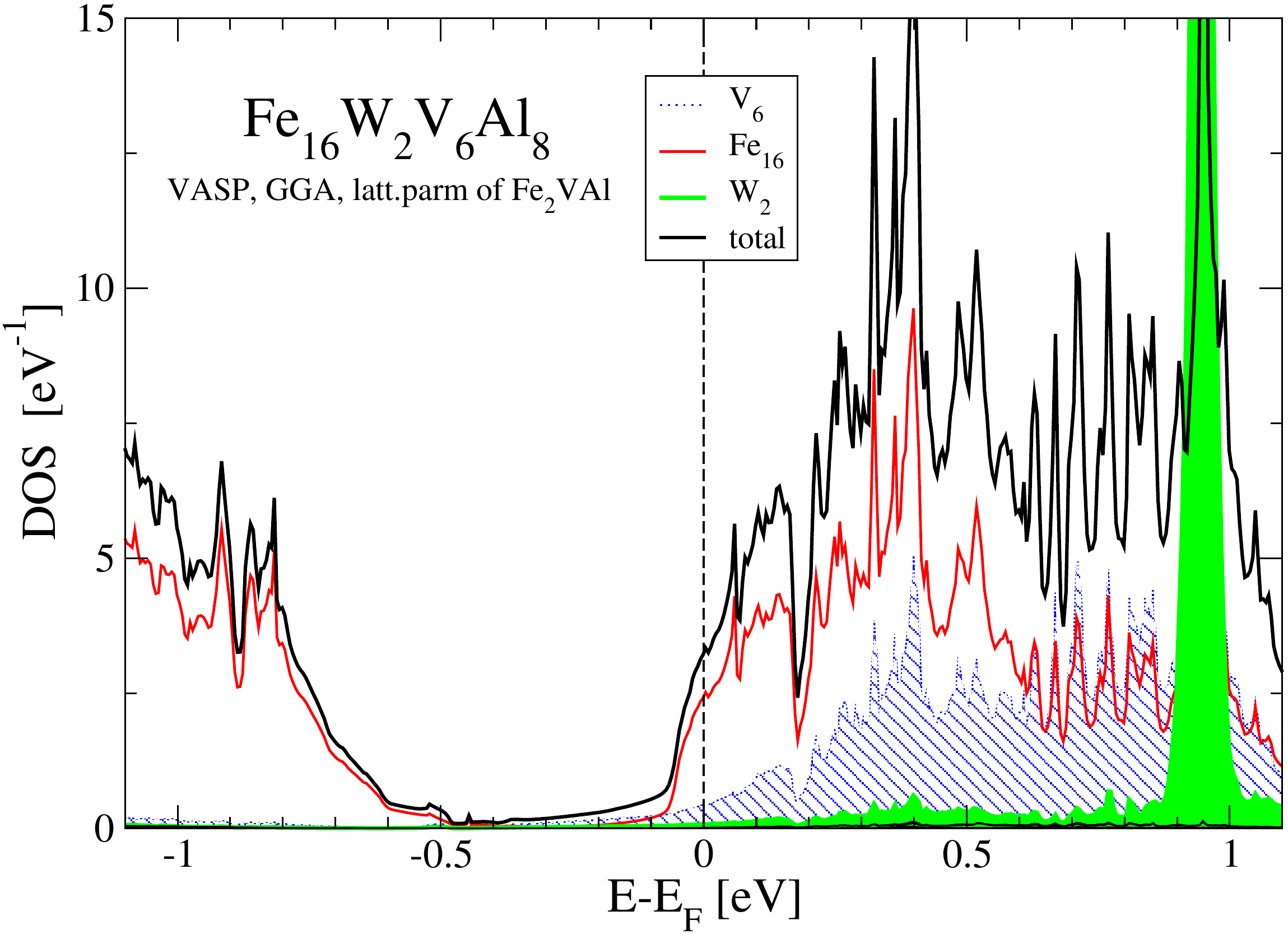}
\end{center}
	\caption{(Color online) Energy dependent partial and total electronic density of states $N(E)$ 
	near to the Fermi energy $E_F$. Upper panel: $\rm Fe_2VAl$; central panel: 
	$\rm Fe_{16}V_7WAl_8$
	(corresponding to $\rm Fe_2V_{0.875}W_{0.125}Al$); lower Panel: $\rm Fe_{16}V_6W_2Al_8$
	(corresponding to $\rm Fe_2V_{0.75}W_{0.25}Al$)
	}
	\label{edos}
\end{figure}

In order to appropriately account for the intensities of the various Bragg peaks, the standard
full-Heusler structure ($\rm Cu_2MnAl$-type, space group no. 225)
was slightly modified such that about 5\,\% of Al is located at the V (4b) 
site and vice-versa, 5\,\% of V is located at the Al (4a) site. It is well known
from literature that different types of anti-side occupation occur in Heusler systems; as
a consequence, a variety of cubic structure types can be realized. Specifics of this
interesting phenomenon are discussed in detail in Ref. \cite{Graf2011}.  
Based on these Rietveld fits, the lattice parameters of the entire 
series were obtained; values are summarised in 
Fig. \ref{xray}, lower panel, together with a sketch of the crystal structure of $\rm Fe_2VAl$. 
Obviously, there is a monotonic increase of the lattice parameter $a$ with increasing W content,
as a result of the larger atomic radius of W (193\,pm) with respect to V (183\,pm)

Results of the first-principles electronic structure calculations 
carried out for $\rm Fe_2V_{1-x}W_xAl$ are summarised in Fig. \ref{edos}.

For stoichiometric $\rm Fe_2VAl$, 
a pseudo-gap in the electronic density of states $N(E)$ is found, 
in full accordance with earlier full potential calculations of Weinert \cite{Weinert}. 
The Fermi energy $E_F$ is located right at the edge of the valence band.
An estimation of the gap width reveals values well below 1~eV.
Below and above the gap, there is a strong energy dependence 
of occupied and unoccupied states (see our previous study on $\rm (Fe,Ni)_2VAl$ \cite{Igor}).

The non-isoelectronic substitution of V by W provides extra electrons to the system; as
a consequence, the Fermi level should rise. Indeed, this can be observed 
from the present DFT calculations carried out 
for the substituted samples in terms of supercells, i.e., 
the chemical composition $\rm Fe_2V_{0.875}W_{0.125}Al$ would 
correspond to $\rm Fe_{16}V_7WAl_8$ and $\rm Fe_2V_{0.75}W_{0.25}Al$
is equivalent to $\rm Fe_{16}V_6W_2Al_8$. Both samples are 
near to real compositions of the present study. In both cases,
$E_F$ moves towards the edge of the conduction band, while 
the pseudo-gap structure is maintained. The colour codes indicate 
that predominantly Fe contributes to the electronic densities in the vicinity 
of the pseudo-gap, while the bunch of W-states is about 1\,eV above $E_F$.
The overall eDOS structure keeps pretty much the same, but 
the densities almost double from $x = 0.125$ to $x =0.25$, i.e., 
$N(E) \approx 1.5$\,states/eV ($x=0.125$), and 
 $N(E) \approx  3$\,states/eV (x=0.25). Moreover, the slopes of $N(E)$
 slightly flatten with increasing $x$. This might distinctly 
 influence the absolute thermopower values. 
The shift of the Fermi energy from the valence band edge in case of $\rm Fe_2VAl$ to the 
conduction band edge for finite W content should go along with a change of the majority charge
carriers, from $p$-type to $n$-type electronic transport. This, in fact, is obvious 
from the present experimental data (see below).

\subsection{Hall data}

In order to qualitatively and quantitatively check the results of electronic structure calculations and
to obtain information about the mobility of the charge carriers, Hall measurements were
carried out and analyzed in terms of a single band model.

\begin{figure}[tbh]
\begin{center}
		\includegraphics[width=0.45\textwidth]{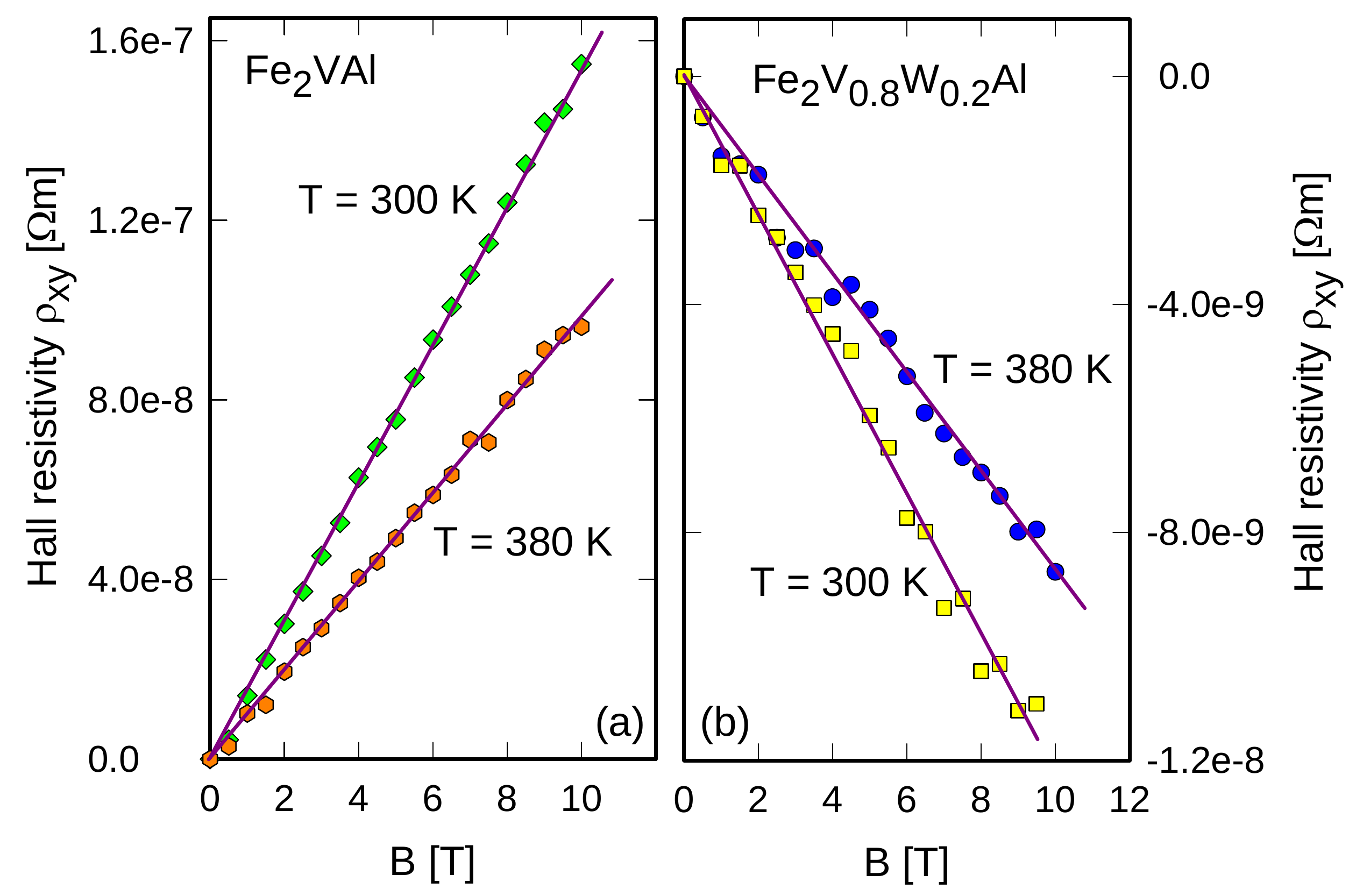}
\end{center}
	\caption{(Color online) Field dependent Hall resistivity $\rho_{xy}$ of
	$\rm Fe_2V_{1-x}W_xAl$ for $x = 0$ (left panel) and $x = 0.2$ (right panel) for various 
	temperatures.
	}
	\label{Hall_1}
\end{figure}

\begin{figure}[tbh]
\begin{center}
		\includegraphics[width=0.45\textwidth]{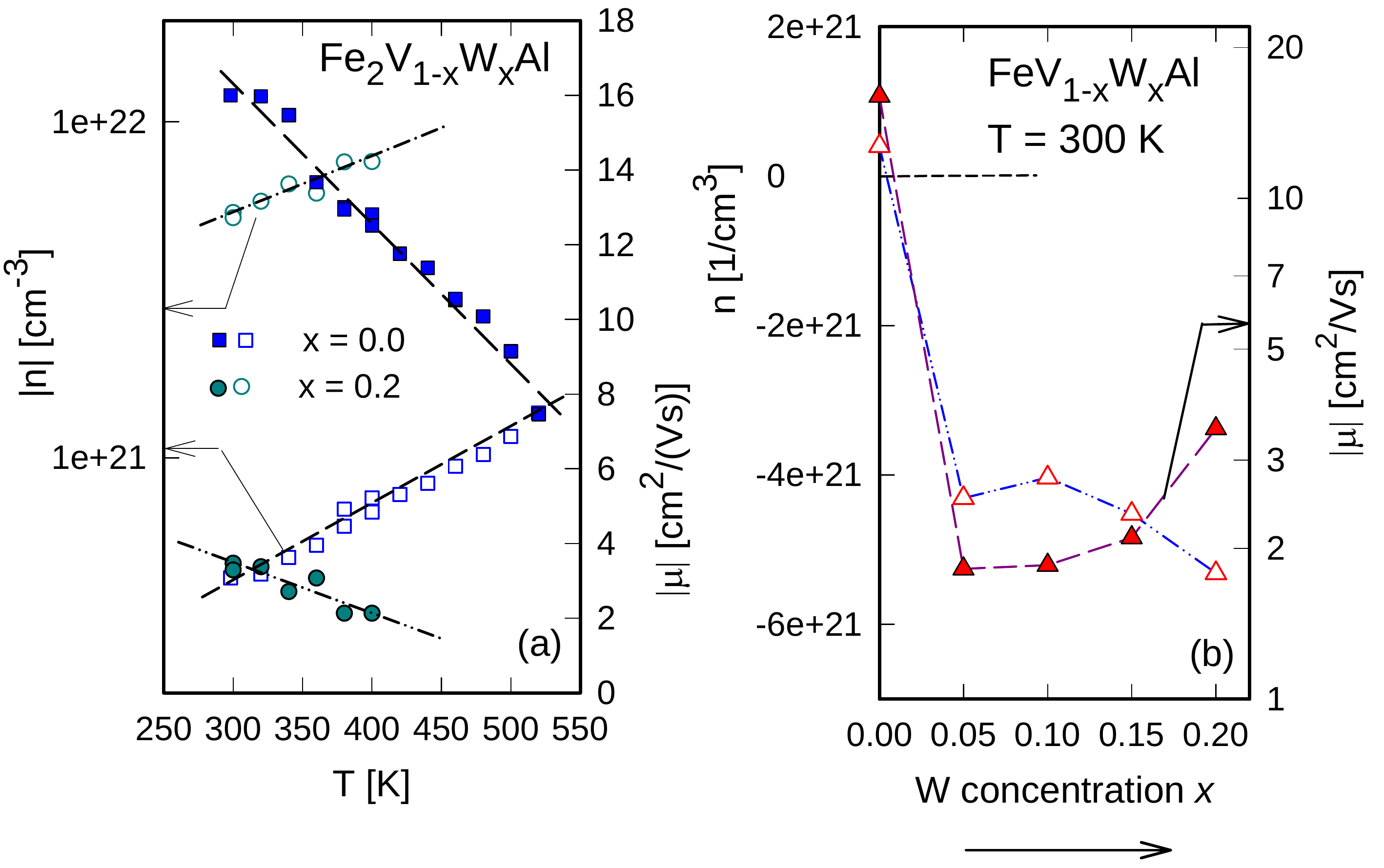}
\end{center}
	\caption{(Color online) (a): Temperature dependent absolute charge carrier density 
	$n$ (left axis) and absolute mobility $\mu$ (right axis) of
	$\rm Fe_2V_{1-x}W_xAl$ for $x = 0$ and $x = 0.2$  (b): Charge carrier density $n$ (left axis) and 
	absolute mobility $\mu$ of $\rm Fe_2V_{1-x}W_xAl$ for various concentrations of $x$, derived
	at room temperature. 
		}
	\label{Hall_2}
\end{figure}

Shown in Fig. \ref{Hall_1} are Hall measurements carried out at $T =300$ and 380\,K
as typical examples of our measurements and fields up
to 10\,T. Obviously, the Hall resistance $\rho_{xy}$ is different for $\rm Fe_2VAl$ and the
W substituted system. While the former is positive and increases linearly with field, the latter
is negative for the entire field range up to 10 T. These observations allow to conclude that
i) there is for both materials a predominating charge carrier type. Thus, all analyses 
can be done in terms a the classical single band model. ii) The positive Hall resistivity 
$\rho _{xy}$ of $\rm Fe_2VAl$ refers to holes as the dominating carriers, while the 
negative one of $\rm Fe_2V_{0.8}W_{0.2}Al$ is indicative of electrons. 
The charge carrier concentration in the single band model follows from
$\rho_{xy} = B / ( \pm q n)$, where $B$ is the magnetic field and q is the unit charge 
of the charge carriers. The mobility $\mu$ is then simply given as 
$\mu = \rho_{xy}/(\rho_{xx} B)$, where here $\rho_{xx}$ represents the ordinary electrical resistivity.
Both, the charge carrier density and the mobility are summarized in Fig. \ref{Hall_2}, left panel,
for temperatures up to 520\,K. 
Obviously, for both samples, the charge carrier concentration grows substantially with increasing
temperature, as a result of excitations of carriers across 
the narrow gap in the density of states. Rising $n(T)$ in conjunction with the observed 
electrical resistivity causes in decrease of the mobility $\mu$, Fig. \ref{Hall_2}, right axis.

Comparing $x = 0$ with the substituted sample $x = 0.2$ reveals that the W substituted 
sample exhibits a charge carrier density $n$, which is almost one order of magnitude larger
than that of $\rm Fe_2VAl$, while the opposite occurs for the mobility $\mu$, which is significantly
lower for the substituted alloy.

The rather small carrier concentration for $x=0$ ($n_h \approx 4 \times 10^{20}$\,cm$^{-3}$) 
at $T = 300$~K convincingly 
corroborates our DFT results and the positive signature refers to the Fermi energy located 
in the tail of the valence band with very small, but finite values of the eDOS. 
In addition, this positive numerical value is in excellent agreement to earlier reports 
deduced from limited field ranges ($B < 1.5$\,T) \cite{Kato_1998,Nishino_2006}.
Because of the fully ordered structure in ternary $\rm Fe_2VAl$, the mobility of 
the holes is high since the number of scattering events is reduced. On the other hand,
even small substitutions of V/W drive the Fermi energy across the pseudo-gap into the 
tail of the conduction band. Electrons are then the main carriers
(compare Fig. \ref{Hall_2}, right panel, $T = 300$\,K),
as also evidenced from the DFT results of Fig. \ref{edos}. In fact, all W substituted 
alloys exhibit negative charge carriers and thus a negative value of the mobility. The charge carrier 
density of the W substituted alloys slightly increases 
with increasing $x$, while the mobility in each case is much smaller than in the starting
material. Disorder due to the V/W substitution might be the driving force behind.

\subsection{Electrical resistivity}

\begin{figure}[tbh]
\begin{center}
		\includegraphics[width=0.45\textwidth]{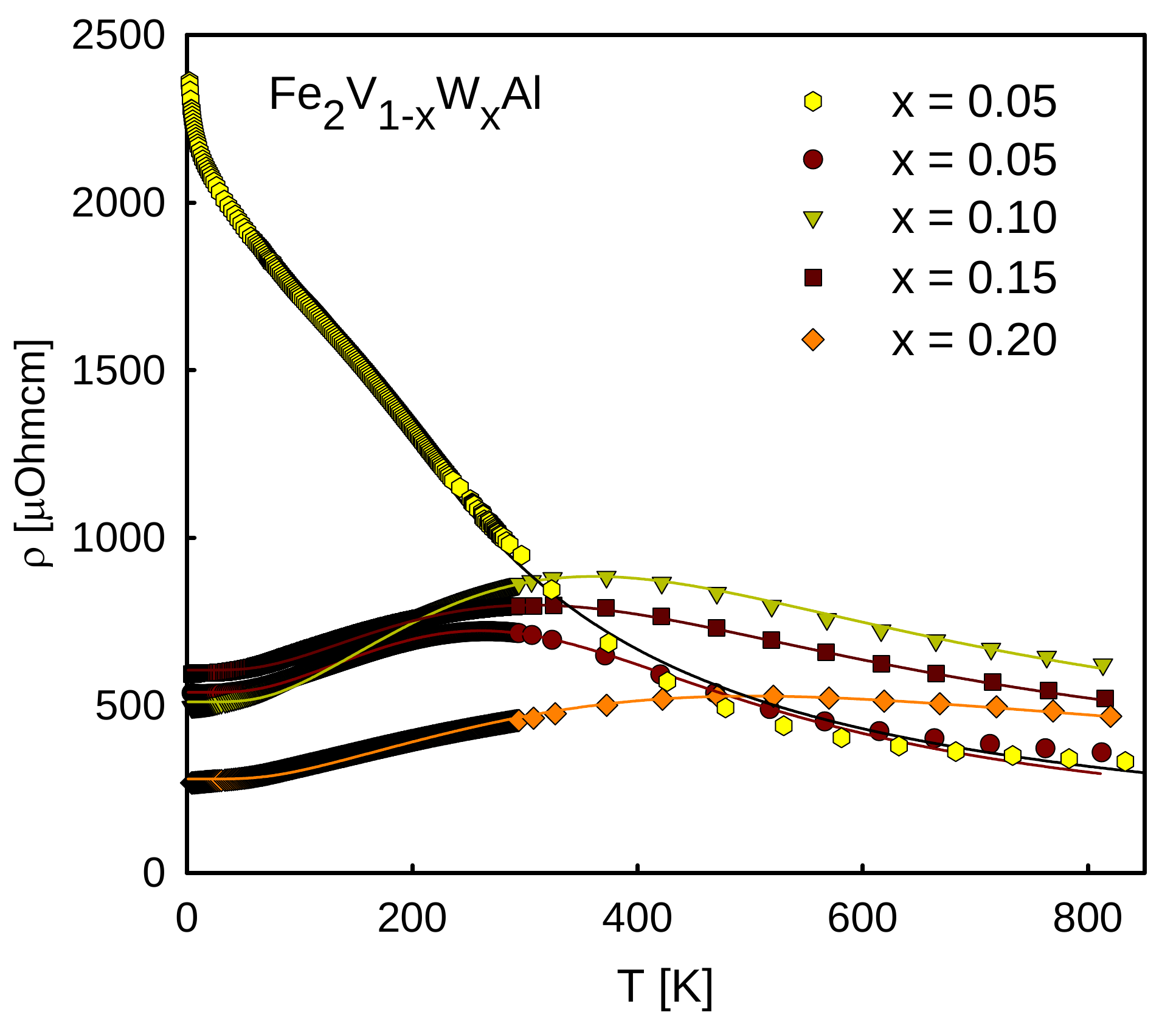}
\end{center}
	\caption{(Color online) Temperature dependent electrical resistivity $\rho$ of  
	$\rm Fe_{2-x}Ni_xVAl$ for various concentrations $x$. The solid lines are least squares 
	fits (see the text).  
	}
	\label{rho}
\end{figure}

Shown in Fig. \ref{rho} is the temperature dependent 
electrical resistivity $\rho$ for 
various concentrations of W in $\rm Fe_2V_{1-x}W_xAl$. 
Obviously, $\rho(T)$ of this series neither follows
a simple metallic behaviour nor is characterized by a typical
semiconducting one. Rather, a subtle overlay of both scenarios is quite
likely; hence the temperature dependent resistivity of the present samples consists of 
metallic like and semiconducting like contributions, together with faint,
sample dependent maxima in $\rho(T)$. The more semiconducting like temperature dependence
of $\rho (T)$ above room temperature coincides with measurements of Mikami et al. \cite{Mikami_2012}.
There, however, no low temperature data were reported.

The distinct behaviour of $\rho(T)$ as present in Fig. \ref{rho},
can be accounted for in terms of a temperature dependent charge carrier
density in the context of  electron-phonon and electron-static imperfection interactions.
A model with a simplified electronic density of states, which we have developed previously 
(for details see e.g., Ref. \cite{Igor}), was successfully applied to various skutterudites,
clathrates and half-Heusler systems, and recently 
also to the Heusler series $\rm Fe_{2-x}Ni_xVAl$ \cite{Igor}.
Least squares fits of the model described above to the experimental data
have been carried out; results are shown as solid lines in Fig. \ref{rho}. 
Over a very broad range of temperatures, the fits reveal convincing 
agreements and a bunch of material related parameters, 
characterizing these thermoelectric materials.

Comparing $\rho(T)$ of various concentrations, several trends are obvious.
The almost semi-conducting like resistivity of starting $\rm Fe_2VAl$ crosses
over to a feature with a smooth maximum in $\rho(T)$ at $T_{\rho}^{max}$. The latter increases
with increasing W concentration. Forming a maximum in $\rho(T)$ is due to the thermal excitation
of electrons across the gap in the eDOS of the system. Qualitatively, this means
that the gap of the series grows with growing W content. 
In fact, the least squares fits reveal an increase of the gap, 
from about 45\,meV for $\rm Fe_2VAl$
to 205\,meV for the sample with $x=0.2$.
The resulting $\rho(T)$ curve 
itself is a subtle balance of the growing charge carrier density, as the temperature increases,
with respect to an increasing electron-phonon interaction with increasing temperature.
Moreover, the increasing charge carrier concentration with increasing W content competes with 
the tendency of the lattice to become more disordered, which would enlarge the residual resistivity
owing to the V/W substitution.

Following the evolution 
of the electronic density of states, the substantially reduced eDOS at $E_F$ slightly below the 
pseudo-gap causes an almost simple semi-conducting behaviour; the temperature dependence,
however, does not coincide with a simple exponential decay. The increase of the W content
causes a change from hole to electron-type charge carriers, together with a density 
increase. This causes, overall, a decrease of the absolute resistivity values and 
a slight shift of the maxima towards higher temperatures. Obviously, since the 
sample with the largest W concentration exhibits the lowest electrical resistivity in the
entire temperature range, the increase in disorder scattering is supposed to be overcompensated 
by an increase of the charge carrier concentration due to the V/W substitution.
The latter is in agreement with results from Hall measurements (compare Fig. \ref{Hall_2}.

\subsection{Thermopower}

\begin{figure}[tbh]
\begin{center}
		\includegraphics[width=0.45\textwidth]{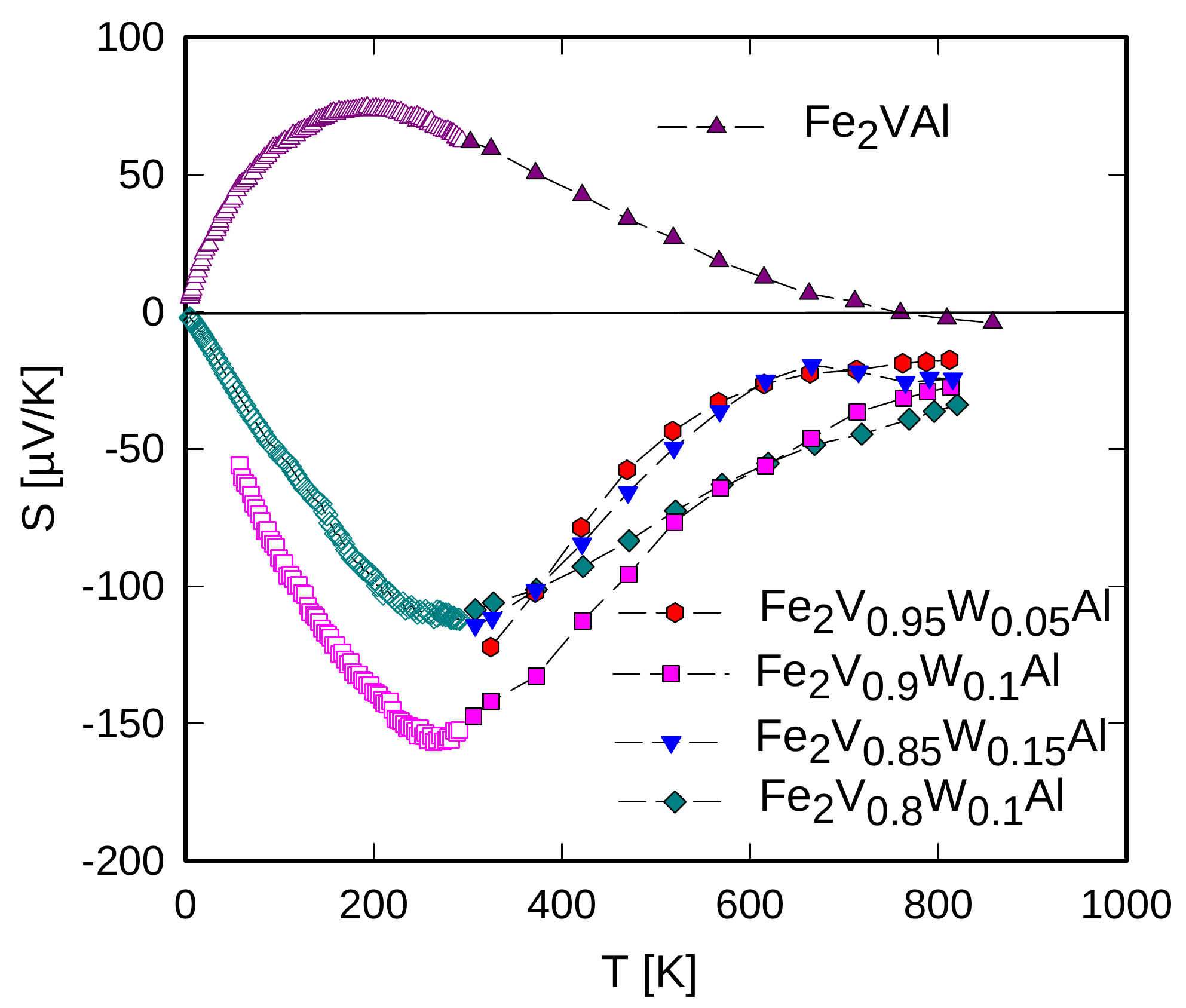}
\end{center}
	\caption{(Color online) Temperature dependent thermopower $S$
	for various concentrations $x$  of 	$\rm Fe_{2}V_{1-x}W_xAl$.
	}
	\label{seebeck}
\end{figure}

The temperature dependent thermopower, $S$, is summarized in Fig. \ref{seebeck} for temperatures 
in a range from 4~K to about 800~K. 
It is interesting to note that the absolute thermopower values, e.g., around 300~K, trace the behaviour
of the charge carrier density as observed from the Hall data (Fig. \ref{Hall_1}), where the smallest $n$-values 
(around $x = 0.1$) reveal the most largest thermopower, about $-150\,\mu$V/K. 
Seebeck values of about the same size and concentration dependence 
for temperatures above room temperature where reported previously by Mikami et al. \cite{Mikami_2012}, corroborating 
the present data, as well as a comparable sample quality of the substituted alloys. 
As already indicated from both the Hall effect as well as from the electronic structure 
calculations, the V/W substitution is responsible for a change of the charge 
carrier type, from hole dominated  ($\rm Fe_2VAl$) to electron dominated 
transport. As a consequence, thermopower changes sign, from positive ($\rm Fe_2VAl$)
to negative (W-substituted samples). This becomes obvious, too, from 
Mott's theory of thermopower,
\begin{equation}
S(T) = C T \frac{1}{N(E)}\frac{\partial N(E)}{\partial E}\vert _ {E = E_F} 
\label{Mott}
\end{equation}
which indicates that the sign and the absolute thermopower values are
determined by the logarithmic derivative of the 
density of states with respect to energy at the Fermi energy. 
In Eqn. \ref{Mott}, $C$ is a constant. 
Taking into account Eqn. \ref{Mott} in the context of the eDOS of the 
present materials (compare Fig. \ref{edos}) 
would provoke larger $S(T)$ values for $x=0.1$ than for 
$x = 0.2$, since $N(E_F) $ is larger in the latter 
and $\partial N (E) / \partial E) \vert _{E=E_F}$
is smaller.

The simple overall behaviour of $S(T)$ as inferred from Eqn. \ref{Mott}, however,  
becomes modified from the fact that the observed thermopower 
not only consists of contributions from a single charge carrier type. 
Rather, because of the relatively narrow 
gap in $N(E)$ next to the Fermi energy, both
electrons and holes are becoming actively involved, i.e.,
\begin{equation}
S = \frac{\sigma_h S_h + \sigma_e S_e}{\sigma_h + \sigma_e}.
\label{see_p_n}
\end{equation}
Here, $\sigma_{h,e}$ and $S_{h,e}$ are the electrical conductivities and the 
Seebeck coefficients of holes and electrons, respectively.  
Eqn. \ref{see_p_n} indicates that the measured thermopower data of 
narrow gap semiconductors might exhibit quite complicate temperature dependences,
where the almost linear temperature dependence as inferred from Eqn. \ref{Mott}
can become dramatically modified from the temperature dependent change
of the charge carrier density $n$ as well as from the temperature
dependent electron - phonon interaction, distinctly influencing the electrical conductivity
of a certain system. Various measurements carried out on doped/substituted $\rm Fe_2VAl$
samples demonstrate a behaviour which is similar to the one of Fig. \ref{seebeck},
including changes from positive to negative values
(compare e.g. Refs. \cite{Renard_2014,Skoug2009,Mikami_2013}).

The Seebeck coefficient of narrow gap semiconductors 
exhibit very frequently extrema in its temperature dependence. 
According to Ref. \cite{Goldsmid_1999}, the 
gap in the eDOS, $E_g$, near to $E_F$ is related to the 
maximum value of the thermopower at that temperature $T_{max}^S$ according to
\begin{equation}
E_g = 2 e \vert S_{max} \vert T_{max}^S.
\label{goldsmid}
\end{equation}
Here, $e$ is the electron charge $S_{max}$ the thermopower at the 
maximum in the $S(T)$ curve at a temperature $T_{max}^S$.
This extremum is a consequence of a bipolar conduction mechanism,
involving thermal excitations of holes and electrons
across the band gap.
The sign change of $S(T)$ of starting material $\rm Fe_2VAl$ 
at elevated temperatures is thus a result of the promotion of 
electrons across the gap in the eDOS near the Fermi energy $E_F$.

Applying Eqn. \ref{goldsmid} to experimental data of Fig. \ref{seebeck}, allows to 
derive a rough estimation of the pseudo-gap, revealing 
14.3, 42 and 31\,meV for for $x = 0.0,\,0.1$ and 0.2, respectively.

\subsection{Phonon dispersion and thermal conductivity}

Measurements of the temperature dependent thermal  
conductivity, $\lambda$, of $\rm Fe_2VAl$ based
alloys reported so far (see e.g., Refs. \onlinecite{Nishino_2006,Lue_2007}), 
demonstrated that the lattice thermal conductivity 
dominates the overall measured quantity. Hence, the knowledge of the phonon system
in such materials is of importance to understand the respective behaviour.
We have carried out DFT calculations in order to obtain 
reliable dispersion and phonon density of state data of these materials.  
Vibrational properties of $\rm Fe_2V_{1-x}W_xAl$ were derived
in the harmonic approximation employing the
direct force-constant-method as implemented in the program package 
\textit{f}PHON.\cite{Alfe2009,DReith} For deriving the force constants, density
functional perturbation theory  calculations as implemented in 
VASP were performed. Atomic positions were optimized until the 
residual forces were less than $2\times10^{-5}$\,eV/\AA{}. For anharmonic 
contributions, a quasi-harmonic approach was applied \cite{Zhang,RMoser}. 
For this approach, bulk moduli are needed which amount to
207 GPa for $\rm Fe_2VAl$ and 218 GPa for $\rm Fe_2V_{0.875}W_{0.125}Al$
and 226 GPa for $\rm Fe_2V_{0.75}W_{0.25}Al$.

\begin{figure}[]
\begin{center}
		\includegraphics[width=0.45\textwidth]{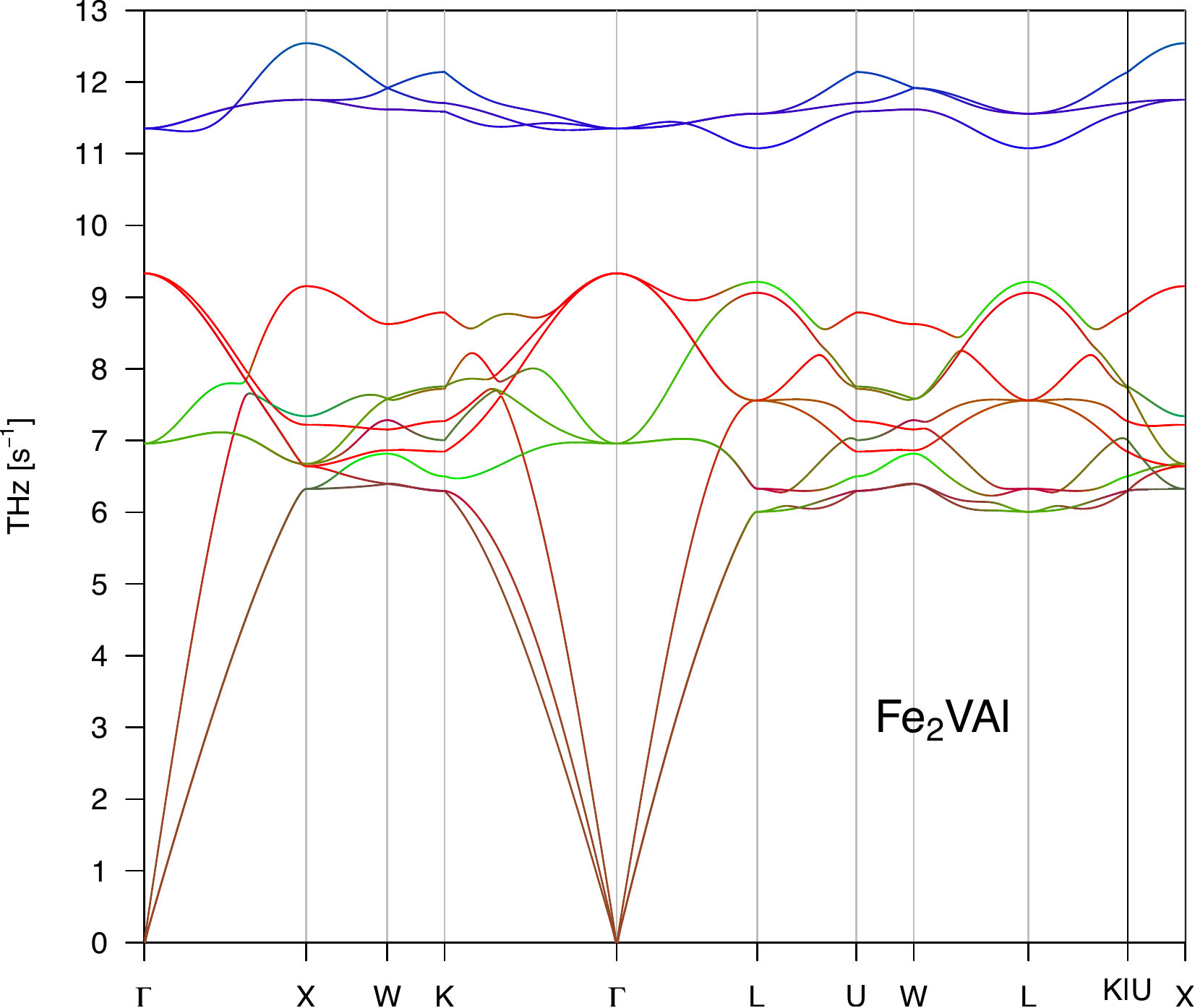}
		\includegraphics[width=0.45\textwidth]{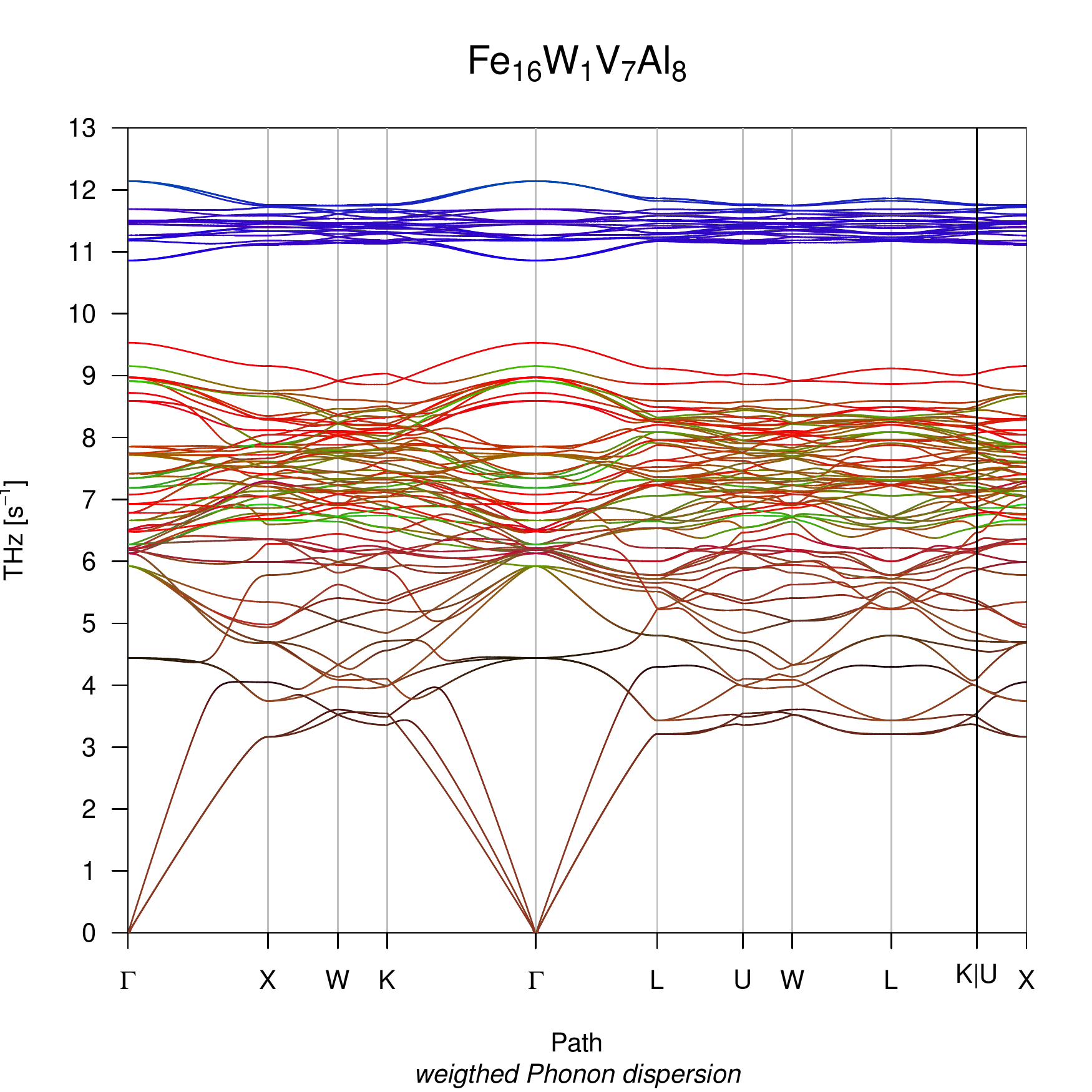}
		\includegraphics[width=0.45\textwidth]{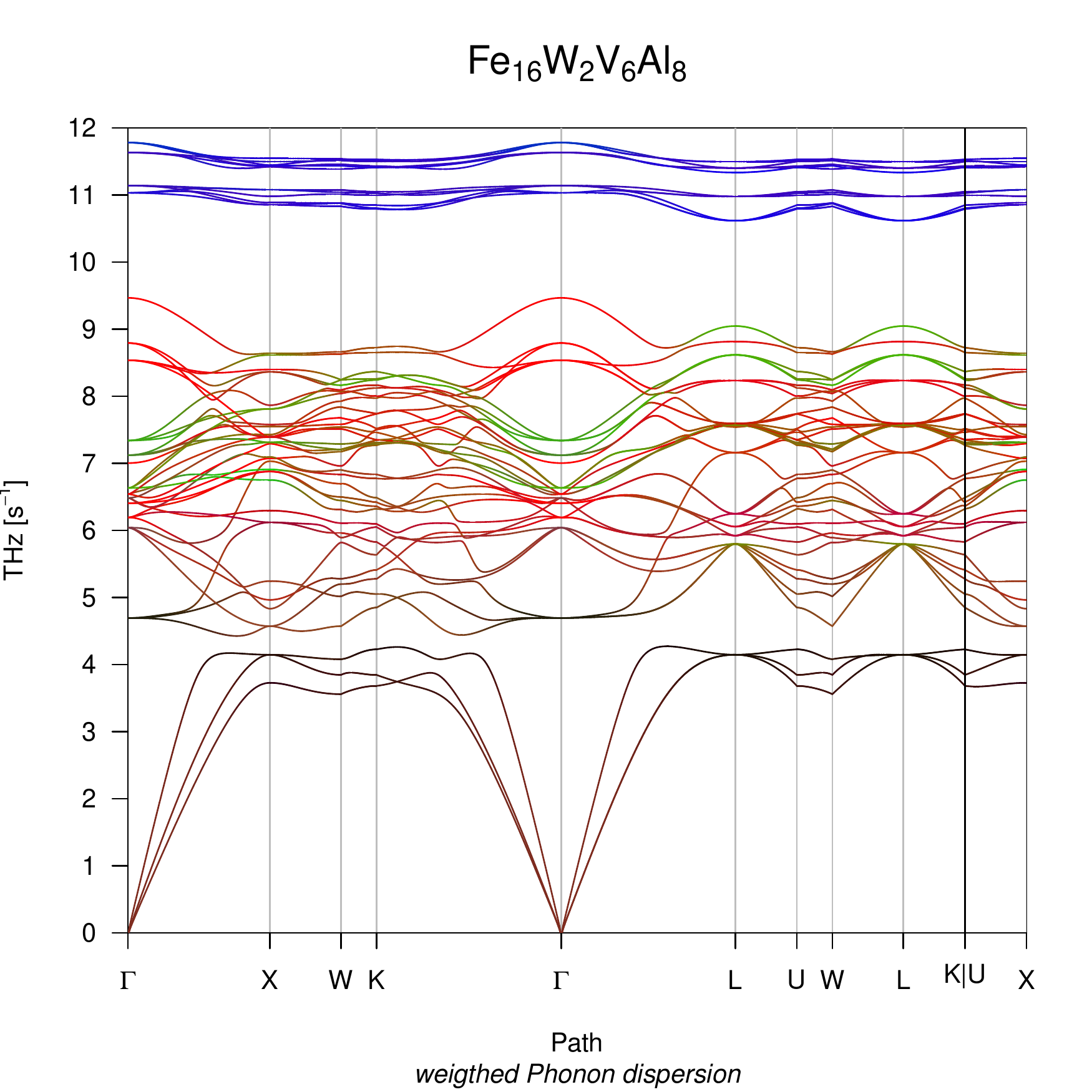}
\end{center}
	\caption{(Color online) Calculated $\vec{q}$ dependent phonon dispersion at DFT
        equilibrium volumes weighted according to local atomic contributions (Fe -- red, V -- green, Al 		
		-- blue, W -- black, mixed colours indicate mixed contributions).
	 	upper panel: $\rm Fe_2VAl$, middle panel: $\rm Fe_2V_{0.875}W{0.125}Al$
	 	lower panel: $\rm Fe_2V_{0.75}W{0.25}Al$.
	}
	\label{p_struct}
\end{figure}

In order to obtain reliable phonon data for 
the W substituted systems, $2 \times 2 \times 2$ supercells have been constructed such that 
$\rm Fe_{16}W_1V_7Al_8$ corresponds to $\rm Fe_2V_{0.875}W_{0.125}Al$
and $\rm Fe_{16}W_2V_6Al_8$ corresponds to $\rm Fe_2V_{0.75}W_{0.25}Al$
Results are presented for the respective DFT volumes in Fig. \ref{p_struct}

The acoustic branches in $\rm Fe_2VAl$ form the quite shallow part 
up to 6~THz, followed by optical branches. Above a gap of about 2~THz, 
almost dispersion-less Al branches are located. 
Relevant parameters with respect to thermal conductivity are the sound velocities 
(2 transversal ($v_{t1,t2}$) and one longitudinal contribution ($v_{l}$)) as well as 
the Debye temperatures. These data are collected in Table \ref{tab2} for $\rm Fe_2VAl$.

Substituting V/W reveals various changes of the overall phonon dispersion (Fig. \ref{p_struct}(b,c)). 
i) The acoustic phonon branches are less steep in comparison to the starting material. As a consequence,
the average sound velocity reduces from 5263 to 4908 and finally to 4632~m/s for $x = 0, \, 0.125, \, 0.25$,
respectively (compare Table \ref{tab2}). 
As the thermal conductivity is a function of the sound velocity, a decrease of $\lambda(T)$ in this
sequence can be expected without considering any further scattering processes. 
Interestingly, the acoustic branches get very flat in the range of about 4~THz, driven mostly 
from W vibrations. ii) The optical branches
are lowered upon the increasing W content, too. A gap opens in the case $x = 0.25$ 
between the acoustic and optical branches and the Al bands lower slightly in both substituted systems. 
Due to the usage of supercells, containing much more atoms, the number of optical branches increases
as a consequence. The lower lying optical bands result from Einstein-like vibrations, which might 
constitute additional scattering of the heat carrying phonons; hence, the thermal conductivity 
of the W substituted alloys is expected to further decrease.
iii) the Gr\"uneisen parameter $\Gamma$ is increasing as well for an increasing W content. 
Since the lattice thermal conductivity $\lambda_{ph}$ depends on 
$\Gamma$, i.e., $\lambda_{ph} \propto 1/ \Gamma^2$,
an additional decrease of thermal conductivity is expected. Note that $\Gamma$ accounts for the 
anharmonicity of lattice effects. 

These three scenarios obtained from our DFT calculations infer that modifications of the 
phonons intrinsically lead to a reduction of the lattice thermal conductivity. Besides, 
enhanced scattering of the heat carrying phonons
on increased static disorder in the lattice due to the random distribution of V and W
on the $(4a)$ sites of the $\rm Cu_2MnAl$-type structure is expected. Primarily mass and sized
differences between V and W matter.

\begin{table}[h]
\caption{Calculated velocities of sound $v_s$  at 300K for DFT equilibrium.
Values given for the  three acoustic modes (l: longitudinal,
t1,t2: transversal) and mode average $\bar{v}_s$, the Debye temperature $\theta_D$ and the
Bulk modulus $B_0$}
\label{tab2}
\begin{tabular}{|l|c|c|c|}
\hline \hline
			    & $\rm Fe_2VAl$ &	$\rm Fe_2V_{0.875}W_{0.125}Al$ & $\rm Fe_2V_{0.75}W_{0.25}Al$ 	\\ \hline
$v_{t1}$ [m/s]	&	4568	&	4247    	& 4013		\\
$v_{t2}$ [m/s]	&	4970	&	4642    & 	4377	\\
$v_{l}$	[m/s]   &   8018	&	7551    & 7234		   \\
$\bar{v_s}[m/s]$&	5263	&	4908    & 4720		 \\
$B_0$ [GPa]	    &	207    &	218   	& 226	 \\
$\theta_D$ [K]	&	696	&	644  	& 606		 \\
\hline \hline
\end{tabular}
\end{table}

The average velocities of sound $\bar{v}_s$ and the 
Debye temperatures were obtained according to
Ref. \onlinecite{ANDERSON}. As discussed previously \cite{Igor},
the DFT related values of $\rm Fe_2VAl$ are larger than those directly derived from 
measurements of the sound velocity, but roughly agree with 
heat capacity studies. The concentration dependent variation of data, however,
are very reasonable. Due to the V/W substitution, the strength of the material
increases, as expressed by the bulk modulus $B_0$, while the increase of mass
upon the increase of the W content reduces the sound velocities and thus the 
Debye temperatures, too, by about 10\,\% for the 20\,\% exchange of V/W.

\begin{figure}[tbh]
\begin{center}
		\includegraphics[width=0.45\textwidth]{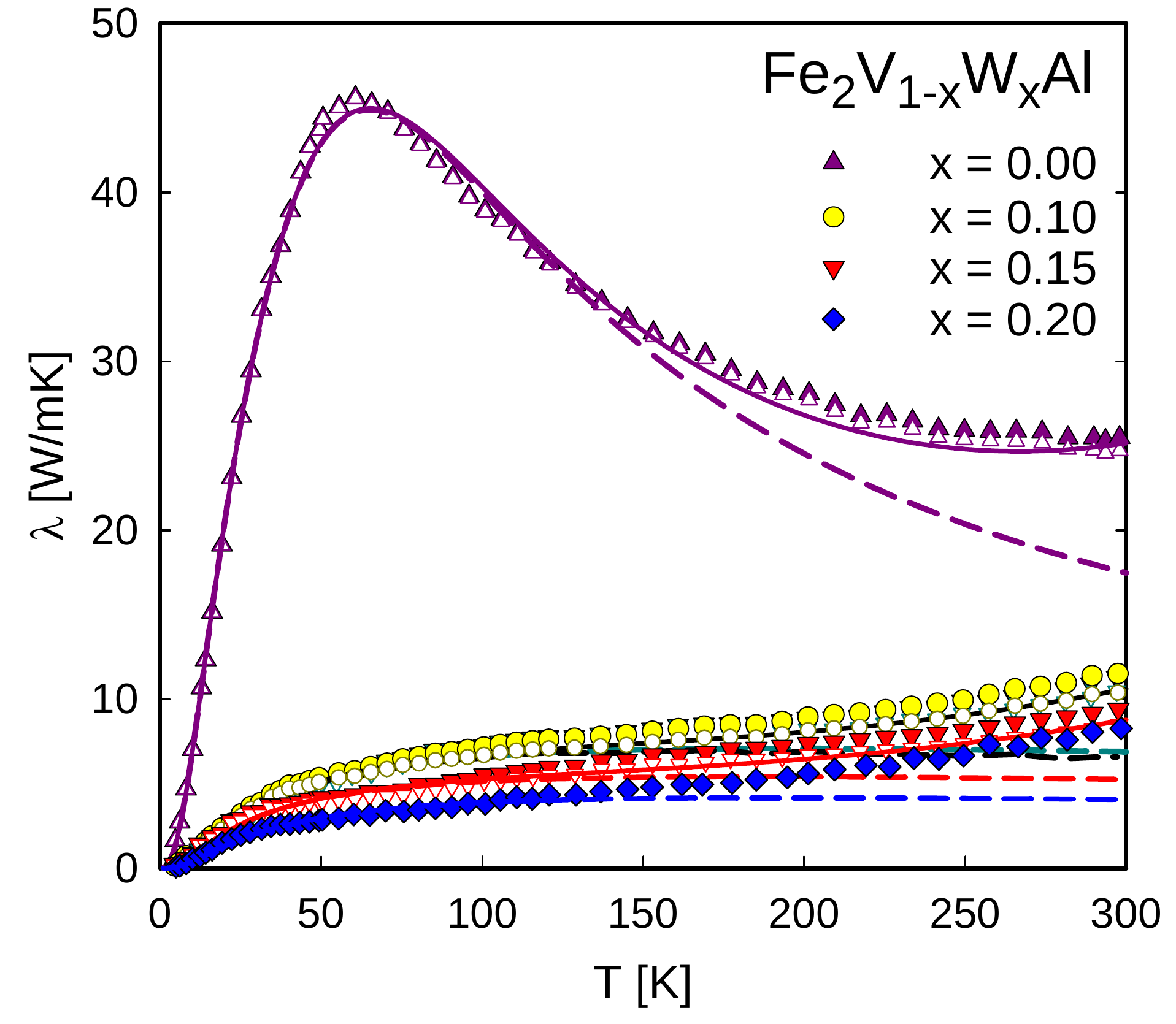}
\end{center}
	\caption{(Color online) Temperature dependent thermal conductivity $\lambda$
	for various concentrations $x$  of 	$\rm Fe_{2}V_{1-x}W_xAl$. The solid and dashed 
	lines are least squares fits as explained in the text.
	}
	\label{lambda}
\end{figure}

The temperature dependent thermal conductivity $\lambda$
of $\rm Fe_{2}V_{1-x}W_xAl$ is plotted in Fig \ref{lambda} 
for various concentrations $x$. It is well known from literature 
that $\lambda (T)$ of $\rm Fe_2VAl$ is, at least in certain temperature ranges, roughly 40 to 50 times
larger than that of archetypal $\rm Bi_2Te_3$. It is, in the most simplest picture, a consequence 
of the very large sound velocity (more than 5200 m/s) and the relative simple type of crystal structure.
As already discussed in the previous paragraph, the V/W substitution causes an intrinsic decrease 
of $\lambda_{ph}(T)$, together with improved scattering on point defects from the disorder at the (4a) sites.
In fact, these mechanisms substantially influence $\lambda (T)$ as can be seen from
the large overall drop of $\lambda (T)$ in the experimental data.

In a standard procedure, as described in many details, e.g., in our recent work on 
$\rm Fe_{2-x}Ni_xVAl$ \cite{Igor}, the thermal conductivity can be 
analyzed and proper conclusions can be drawn. The lattice
thermal conductivity follows simply from $\lambda_{ph} = \lambda - \lambda_e$, 
where the subscript $e$ refers to the electronic thermal conductivity.
The latter is derived employing the Wiedeman Franz law. Obviously, $\lambda_e$
is small; its contribution is just a few percent to the overall measured effect. 
The analysis of the remaining part, $\lambda - \lambda_e$, is carried out in the context
of Callaway's theory \cite{Callaway_1959}, where phonons are treated in terms of the Debye model, and 
scattering processes are accounted for by a total relaxation time, $1/\tau_c = \sum 1/ \tau_i$.
The individual scattering processes are considered independently from each another
and are characterized by distinct temperature and frequency dependencies \cite{Callaway_1959}.

The Callaway model can be expressed as \cite{Zhou_2005}
\begin{equation}
\lambda_{ph} = \frac{k_B}{2 \pi^2 v_s}\left(\frac{k_B T}{\hbar} \right)^3
\int_0^{\theta_D/T} \frac{x^4 \exp(x)}{\tau_c^{-1} (\exp(x)-1)^2}dx .
\label{Callaway}
\end{equation}
$x = \hbar \omega /(k_B T)$, $k_B$ is the Boltzmann 
constant and $\omega$ the phonon frequency; $\hbar = h/(2 \pi)$. 
$\tau_c$ for the relevant phonon scattering processes is given by
\begin{equation}
1/\tau_c = v_s/L + A \omega^4 + B \omega^2 T \exp(-\theta_D/T) + E \omega
\label{tauc}
\end{equation}
where the terms on the rhs of Eqn. \ref{tauc} refer to scattering on grain boundaries with average size $L$,
on point defects, Umklapp-processes and electrons, respectively. $A,\,B$ and $E$ are material dependent 
constants. Least squares fits according 
to Eqns. \ref{Callaway} and \ref{tauc} have been performed and results are
shown in Fig. \ref{lambda} as solid and dashed lines. 
Adding a $T^3$ term to account for radiation losses \cite{Pope_2001}, inherent to the 
steady state heat flow technique,  
reveals a fairly well agreement between experiment and theory.
In all cases, the sound velocity and the Debye temperature 
was changed linearly from $ x = 0 \to x = 0.20$ as inferred from the DFT results.

\begin{figure}[tbh]
\begin{center}
		\includegraphics[width=0.45\textwidth]{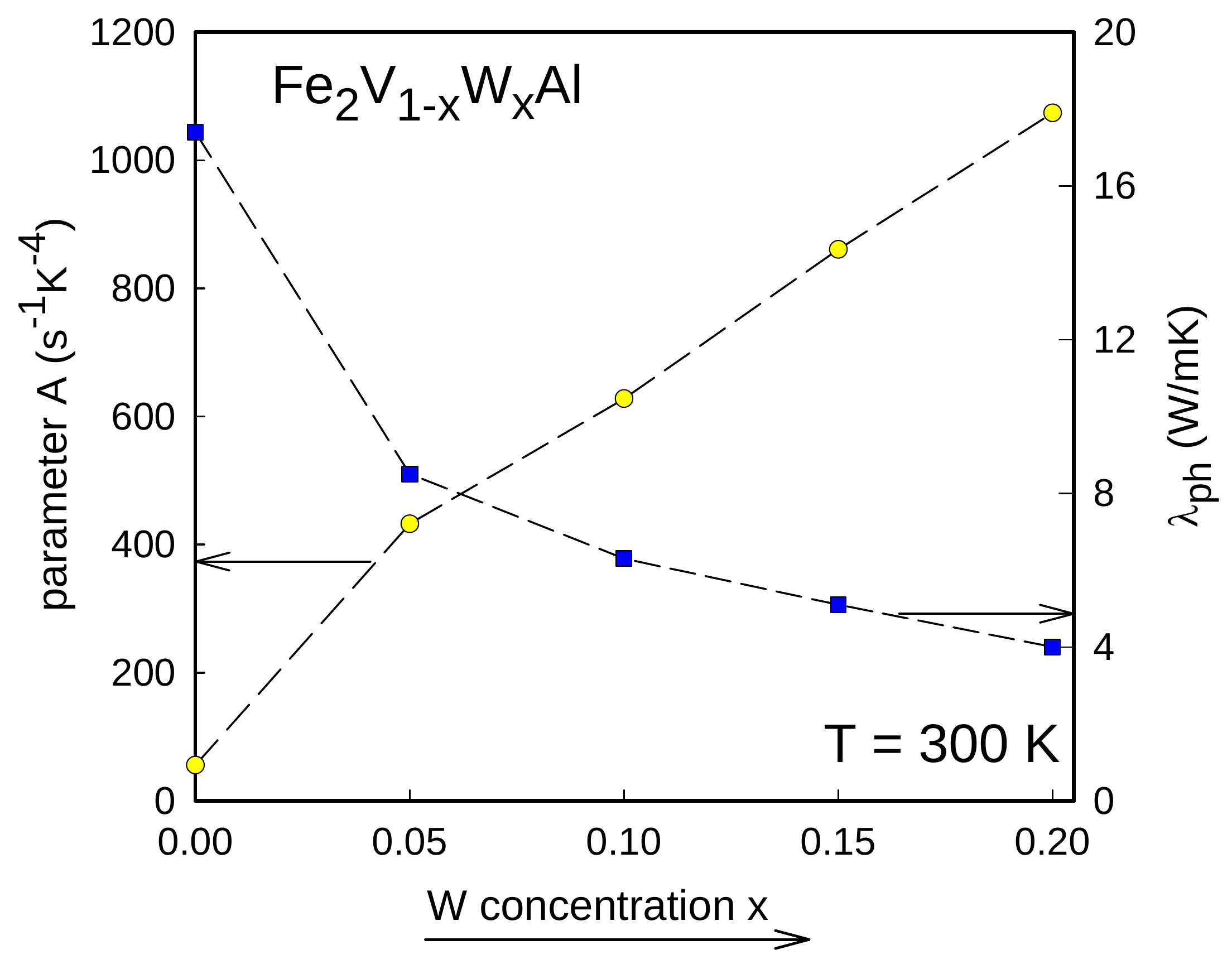}
\end{center}
	\caption{(Color online)  Concentration dependent point defect scattering parameter $A$ 
	of $\rm Fe_{2-x}Ni_xVAl$ (compare Eqn. \ref{tauc}) (left axis) and lattice thermal 
	conductivity $\lambda_{ph}$ taken at room temperature.
	}
	\label{pointdefect}
\end{figure}

The parameters derived from these least squares fits allow to trace, among others, the evolution of the 
point defect scattering parameter $A$ as a function of the W concentration 
(compare Fig. \ref{pointdefect}, left axis). 
It demonstrates that W constitutes a significant scattering center, as this parameter 
grows substantially, by more than one order of magnitude. 
This is about twice as large as observed in a similar study concerning the Fe/Ni substitution in 
$\rm (Fe,Ni)_2VAl$ \cite{Igor} and can be attributed to much larger mass/volume differences between 
V and W compared to Fe and Ni. Because of this significant enhancement of scattering on point defects,
as well as intrinsic parameters like the sound velocity, the thermal conductivity drops, to below 
a quarter of the value of $\rm Fe_2VAl$ at $T=300$~K for the system with 20\,\% V exchanged by 
W (compare Fig. \ref{pointdefect}, right axis). These values deduced at room temperature 
are in almost perfect agreement with the recent study of Mikami et al. \cite{Mikami_2012}, where 
$\lambda_{ph}$ of the sample $x=0.1$ reaches about 4\,W/(m$\cdot$K).

The efficiency of point defect scattering is a result of contributions
from mass fluctuations and volume differences of atoms 
on certain sublattices of the crystal structure.
In general, it can be  expressed 
by a disorder parameter $\Gamma$ as 
$\Gamma = \Gamma_{M} + \Gamma_{S}$, where the subscripts $M$ and $S$
refer to mass and strain field, respectively \cite{Abeles_1963}.

\begin{figure}[tbh]
\begin{center}
		\includegraphics[width=0.45\textwidth]{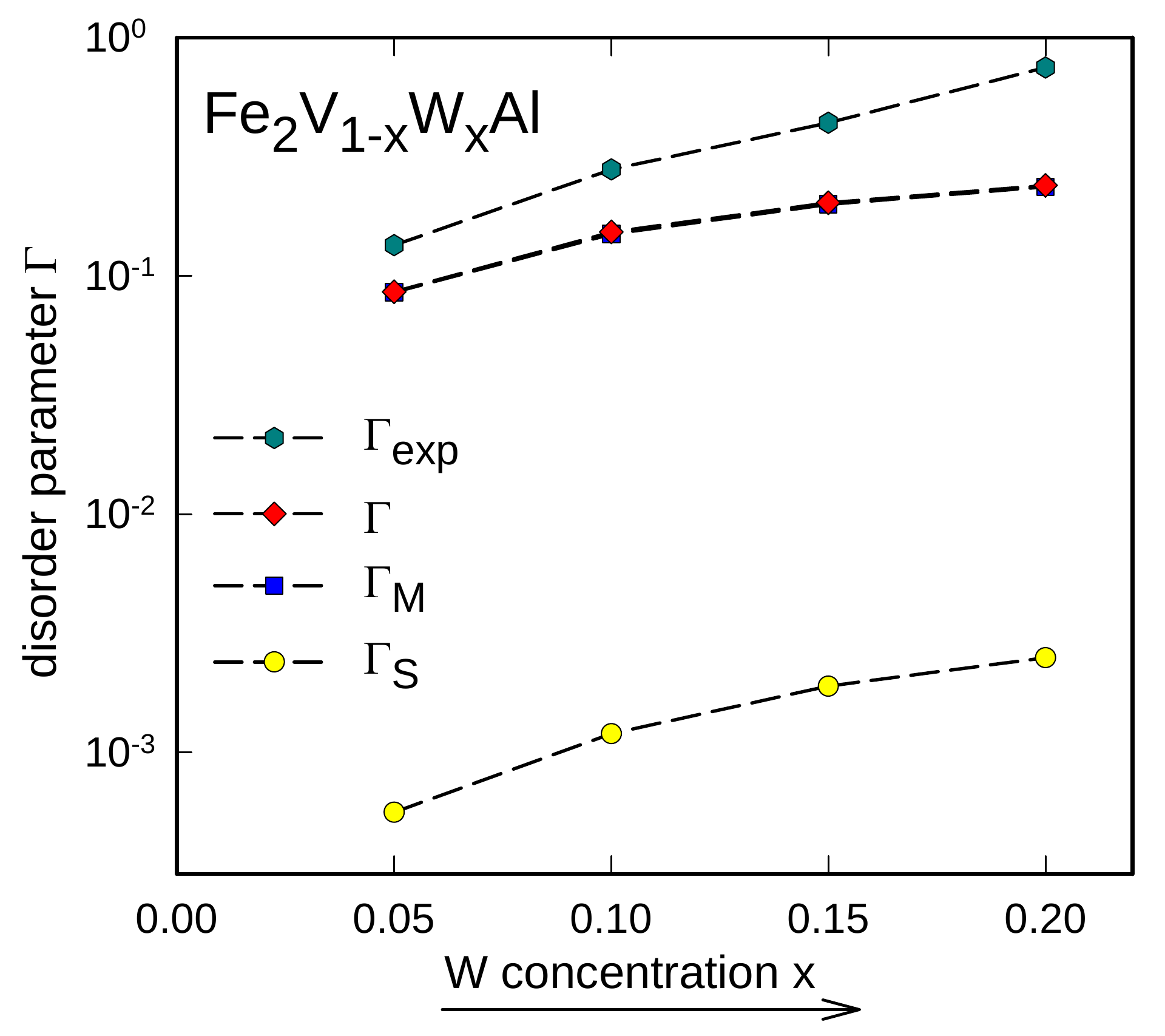}
\end{center}
	\caption{(Color online) Concentration dependent disorder parameter of $\rm Fe_{2-x}Ni_xVAl$
	derived experimentally ($\Gamma_{exp}$) and 
	theoretically ($\Gamma$); the latter consists of an mass ($\Gamma_M$) and volume related 
	contribution ($\Gamma_S$). 
	}
	\label{disorder}
\end{figure}

Following the procedure as outlined in Refs. \cite{Abeles_1963,Igor} reveals
a substantial contribution from mass differences between V and W, almost 2 orders
of magnitude larger than the respective contribution due to size differences (Fig. \ref{disorder}).
The parameter $\Gamma_{exp}$ is derived from the experimental thermal conductivity data,
as described in detail by  Yang et al. \cite{Yang_2004}. The gap between $\Gamma_{exp}$
and $\Gamma$ is closed by an empirical factor (in general between 10 and 100) \cite{Abeles_1963},
accounting for an appropriate pre-factor of the strain field contribution. 
Mikami et al have analysed their data in a similar manner and found that the strain-field contribution
is twice as large as the mass contribution \cite{Mikami_2012}. 
This, however, seems not to be meaningful, since
the difference of the metallic radii between W and V is smaller than 3\,\%, while W has a more than 3 times larger 
mass than V. It is thus very likely that the predominant contribution is due to the mass difference between V and
W as derived in the present analysis (compare Fig. \ref{pointdefect}).

\subsection{Thermoelectric performance}

The relatively large Seebeck values and moderate electrical resistivities
observed in $\rm Fe_2V_{1-x}W_xAl$ samples cause that the power factors
$p_f$ of the four sample studied are ranging between 1.6 and 2.55 mW/(m$\cdot$K$^2$). These 
values are large and compare with those of non-optimized $\rm Bi_2Te_3$ systems. 
The figure of merit, $ZT$, keeps modest because of the still large thermal conductivities
but reaches $ZT \approx 0.20$ for $x = 0.1$.
While $\rm Bi_2Te_3$ based materials have thermal conductivities, ranging between
1 and 4\,W/(m$\cdot$K), the room temperature value found for $\rm Fe_2V_{0.8}W_{0.2}Al$
is of the order of 4\,W/(m$\cdot$K). The reduction of $\lambda(300)$K to about 25~\% of its 
initial value in $\rm Fe_2VAl$ is besides intrinsic changes of 
the phonon system a result of point defect scattering enhanced by the V by W exchange.

\begin{figure}[tbh]
\begin{center}
		\includegraphics[width=0.45\textwidth]{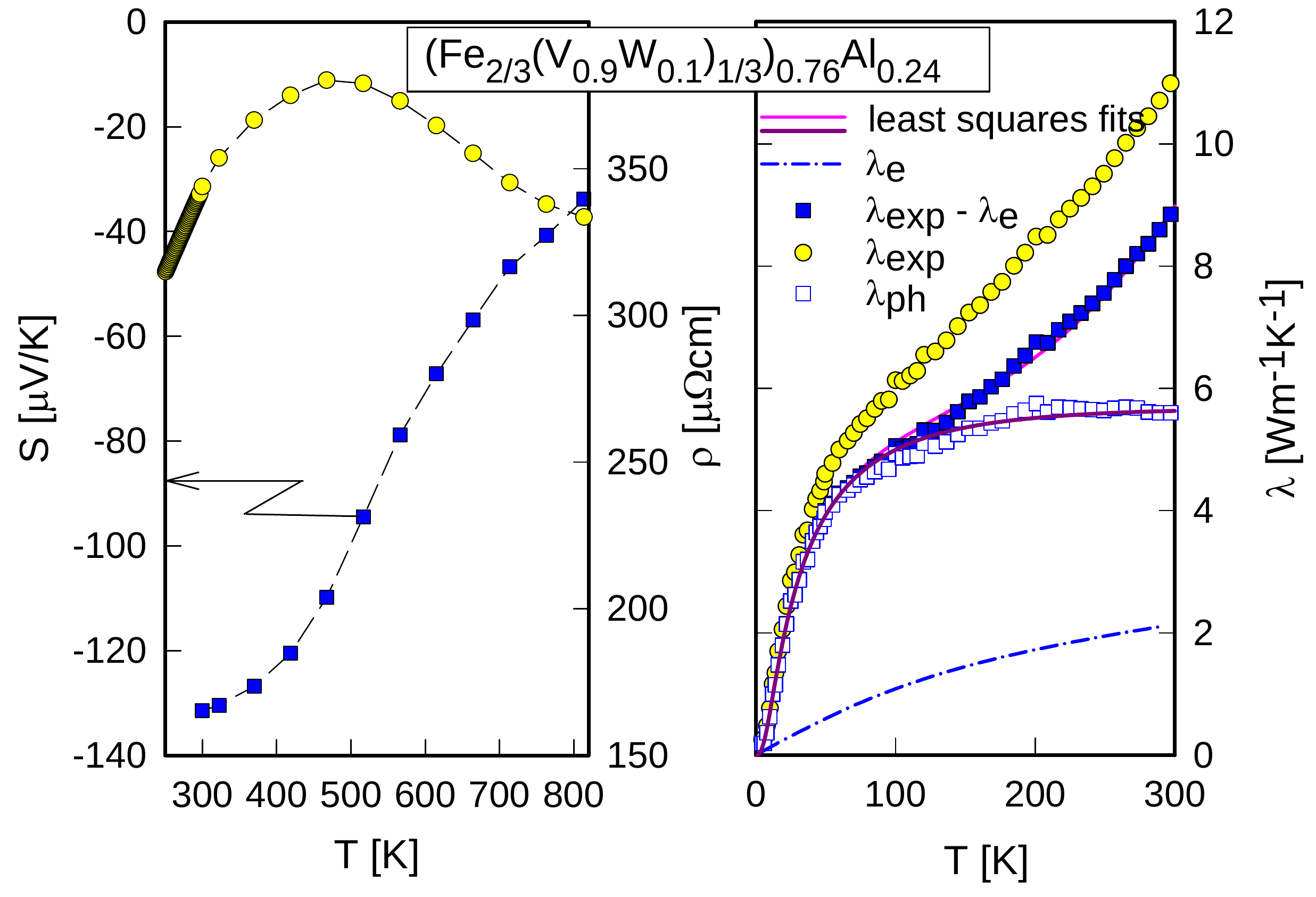}
\end{center}
	\caption{(Color online) Left panel: Temperature dependent 
	Seebeck coefficient, $S$ (left axis) and temperature dependent electrical resistivity, $\rho$,
	of $\rm (Fe_{2/3}(V_{0.9}W_{0.1})_{1/3})_{0.76}Al_{0.24}$. Right panel:
	Temperature dependent 	thermal conductivity, $\lambda$, of
	of $\rm (Fe_{2/3}(V_{0.9}W_{0.1})_{1/3})_{0.76}Al_{0.24}$. The electronic and 
	lattice thermal conductivity are added, together with least squares fits according to 
	Eqns. 4 and 5. 
	}
	\label{off_2}
\end{figure}

\subsection{Off-stoichiometric W-based Heusler systems}

\begin{figure}[tbh]
\begin{center}
		\includegraphics[width=0.45\textwidth]{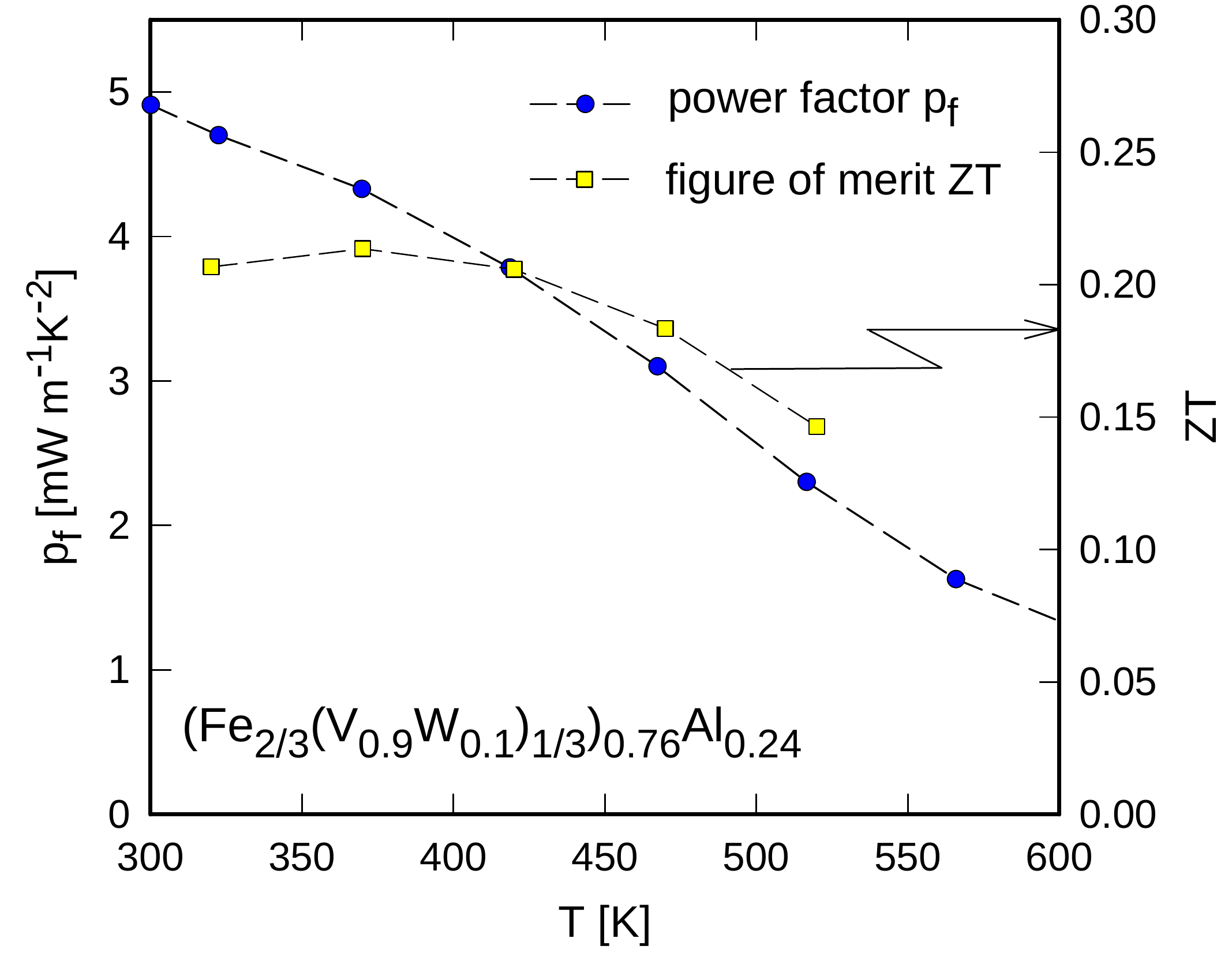}
\end{center}
	\caption{(Color online) Temperature dependent 
	power factor, $p_f$ (left axis) and figure of merit, $ZT$,
	of $\rm (Fe_{2/3}(V_{0.9}W_{0.1})_{1/3})_{0.76}Al_{0.24}$. 
	}
	\label{off_1}
\end{figure}

Besides substitution and doping of a certain starting material, e.g., $\rm Fe_2VAl$,
off-stoichiometric preparation, i.e., $\rm Fe_2V_{1 \pm x}Al_{1 \mp x}$ was already
employed and distinct changes in the electronic and thermal transport were 
obtained. Here, we combine the substitution of V by W with off-stoichiometric
preparation. As an example, $\rm (Fe_{2/3}(V_{0.9}W_{0.1})_{1/3})_{0.76}Al_{0.24}$ is chosen. 
Experimental results are shown in Figs. \ref{off_2} and \ref{off_1}. 
Remarkably, the overall electrical resistivity, $\rho$, is small, 
only about half of the respective stoichiometric compound.
Its temperature dependence is, again, a combination of 
metallic and semiconducting-like contributions. In addition, 
the charge carrier density $n_e = 4 \cdot 10^{21}$~cm$^{-3}$ is almost equal.
This refers to a substantially higher mobility of electrons in this sample. 
The Seebeck effect around room temperature, on the other hand, is large, about -135~$\mu$V/K, indicating 
electrons as primary charge carriers. 
The temperature dependent thermal conductivity
is similar to the equivalent stoichiometric system and the phonon contribution $\lambda_{ph}$
at room temperature coincides within about 10~\%. As a consequence, the power factor, $p_f$,
(compare Fig. \ref{off_1}) appears to be extraordinarily large, exceeding best behaving 
$\rm Bi_2Te_3$ at ambient pressure \cite{Ovsyannikov}. In the context of the reduced thermal 
conductivity observed, the figure of merit, $ZT = 0.22$, is 
one of the largest ever obtained $ZT$ value of full Heusler systems.
Here, it should be noted that while the under-stoichiometric systems (i.e., $\rm Al_{24}$) are electron 
dominated, the over-stoichiometric systems, i.e.,  $\rm Al_{26}$ are hole dominated. This indicates
that small changes of the element contents in $\rm Fe_2VAl$ based Heusler systems might have dramatic effects 
on the electronic structure and thus on electronic and thermal transport.

\section{Summary}

The substitution of V/W in the full Heusler series $\rm Fe_{2}V_{1-x}W_xAl$ is responsible 
for a distinct change of electronic and thermal transport. This is caused by the movement
of the Fermi energy from the valence band edge in $\rm Fe_2VAl$ 
across the band gap towards the conduction band, as proven from Hall and Seebeck effect 
measurements and confirmed by DFT calculations of the electronic structure (band dispersion 
and electronic density of states, eDOS). Since the Fermi energy is located 
at steep slopes of the eDOS, large Seebeck values are expected, and in fact are
found experimentally.

Besides modifications of the electronic structure, the phonon system of the 
the substituted samples changes as well. This results primarily in a reduction of the
various values of sound velocities, a decrease of the Debye temperature and an increase of 
the Gr\"uneisen parameter. These changes drive an intrinsic decrease of the overall thermal
conductivity. However, point defect scattering, mainly because of the lattice
disorder created by the statistical distribution of V and W on the (4a) site of the 
$\rm Cu_2MnAl$ structure reveals an additional contribution to the reduction 
of $\lambda (T)$ because of large mass and volume differences between V and W.
As a result thermal conductivity for the sample $x=0.2$ drops to about 25~\% of its initial value
deduced for $x = 0$. This is a much stronger reduction (almost a factor of 2) compared to a similar 
substitution study of Fe/Ni in $\rm Fe_2VAl$ \cite{Igor}.

\acknowledgments
Research supported by the Christian Doppler Laboratory for Thermoelectricity
and the JST, CREST (JPMJCR19Q4). DFT and related calculations were
performed on the Vienna Scientific Cluster VSC3.

\bibliography{heusler_1}

\end{document}